\documentclass[11pt]{article}
 				\def\rM{{\rm M}}

\def\etal{{\it et al.~}}

\def\half  	{\textstyle {1 \over 2} \displaystyle}

\def\inv16  	{\textstyle {1 \over 16} \displaystyle}
\def\3ov4  	{\textstyle {3 \over 4} \displaystyle}
\def\4ov3  	{\textstyle {4 \over 3} \displaystyle}
\def\8ov11  	{\textstyle {8 \over 11} \displaystyle}
\def\15ov16  	{\textstyle {15 \over 16} \displaystyle}

\def\beq{\begin{equation}} \def\eeq{\end{equation}}
\def\bea{\begin{eqnarray}} \def\eea{\end{eqnarray}}

\def\Tan{{\rm T}}
\def\Ad{{\rm Ad}}
\def\ad{{\rm ad}}
\def\int{{\rm int}}
\def\Aut{{\rm Aut}}
\def\End{{\rm End}}

\def\cC{{\cal C}}

\def\cA{{\cal A}}
\def\cD{{\cal D}}
\def\cDM{{\cal D}_{\rm M}}

\def\cF{{\cal F}}
\def\cG{{\cal G}}
\def\cH{{\cal H}}
\def\cI{{\cal I}}
\def\cJ{{\cal J}}
\def\cK{{\cal K}}
\def\cL{{\cal L}}
\def\cM{{\cal M}}
\def\cN{{\cal N}}
\def\cO{{\cal O}}
\def\cQ{{\cal Q}} \def\cR{{\cal R}}
\def\cS{{\cal S}}

\def\cU{{\cal U}} 

\def\cZ{{\cal Z}}
\def\cSO{{\cal SO}}

\def\Mink{{R$^{1,3}$}}
\def\sdprod {\,{\supset \!\!\!\!\!\!\! \times}\,} 	

\def\Aut{{\rm Aut}} \def\Out{{\rm Out}} \def\Inn{{\rm Inn}} 
\def\Id{{\rm Id}}  \def\Ker{{\rm Ker}}    
\def\lalpDer{ {_\alpha{\stackrel{\rightarrow}{\cD}}}} 			
\def\rD{ {\stackrel{\leftarrow}{\cD}}} 
 	 	
\def\cOut{{{\cO}ut}}
\def\OutlalpDer{ {^{_\cOut}_\alpha{\stackrel{\rightarrow}{\cD}}}}

\begin{document}

\title{
On the Geometry of Spacetime I: \\ baby steps in quantum ring theory
}

\author {Rafael A.
Araya-G\'ochez\footnote{
  	{\small Physics Department, Occidental College M21,}
	{\small 1600 Campus Rd., Los Angeles, CA 90041} 
  	}
}


\date{}
\maketitle

\begin{abstract} 	
 Vierbeins provide a bridge between the curved space of general relativity and the flat tangent space of special relativity.  Both spaces should be causal and spin. 
 We posit intertwining the two symmetries of spacetime bundles {\it asymmetrically}; 
 disentangling the non-trivial {\bf Id} between the base, curved space as a locally ringed space and its Zariski (co-)tangent space.
 This involves the introduction of a ``two-sided vector space" as a section of the smooth, stratified diffeomorphism bundle of spacetime.
 A change of paradigm from the fiber bundle approach ensues where the bundle space takes an active role and the group actions are implemented through asymmetric
 ``scalar multiplication" by elements of a skewed K-algebra on a free K-bimodule.
 Accordingly, the left action is augmented from that on the right {\it algebraically} by a left-sided algebra automorphism
 {\it via} a left $\alpha$-derivation as a non-central Ore extension of a Weyl algebra.
 Curiously, summoning the left $\alpha$-derivation in the context of spacetime symmetries may constitute the key to an {\it asymmetric quantization} of the theory.
 Furthermore, it is conjectured that causal and spin structure may be endowed upon the spacetime itself, independently of the tangent space structure. 
\end{abstract}    


\pagebreak
\tableofcontents
\pagebreak 		

\section{Introduction}
 \label{sec:Intro}

\begin{quote}
{\footnotesize \it
One advantage of a runaway rather than a true cosmological ``constant" is that, by analogy with a zero cosmological constant scenario first outlined by Dyson 
many years ago, in a runaway scenario life can possibly adapt and survive and develop forever by working at lower and lower temperatures and with longer time 
and length scales.  A strict cosmological constant would bring this to a grim end by introducing effective length and time cutoffs.  
\begin{flushright}
Ed Witten, Dark Matter 2000, Marina del Rey, CA.
\end{flushright}
}
\end{quote}

	Nearly 140 years ago, Felix Klein was first to notice that certain invariant geometrical properties--namely angles, parallelisms, and cross-ratios--gradually
 loose invariance as the transformation group that embodies the geometry is enlarged.
 Such an insight led to his celebrated {\it Erlangen Programme}: the characterization of classical geometries on the basis of the transformation groups that 
preserve intrinsic geometrical invariants\cite{Klein1872}.  
 In modern parlance, the latter group is presented as the group of automorphisms of the geometry thus embodying the notion of ``symmetry". 
 Furthermore, the following hierarchy of symmetry groups: Euclidean $\subset$ Affine $\subset$ Projective, naturally led to the notion of homogeneous model geometry 
 as a coset space: the quotient of a transitive Lie group epitomizing the geometry by the stabilizer sub-group of a point in the manifold. 
 Implicit in this geometrical construction is that the manifold  
 and the group action both be smooth.  Important modern generalizations of the notion of Lie group structure include topological manifold with continuous action
 and algebraic variety with regular action under the Zariski topology. 	

	Notably, a homogeneous model space need not admit Lie group structure unless the quotient is by a normal sub-group of the original group.
 Common knowledge holds that the product of equivalent classes in a coset space with group structure does {\it not} depend on the choice of ``representative" 
group elements\footnote{\label{FN:NonSubG}
Groupwise, given a closed subgroup H $\subset$ G, define the {\it left} coset (or quotient) space as the set of all equivalent classes:
G/H = $\Sigma_{g_i \not\in g_{i-1} {\rm H} } ~[g_i]$ 
where $g_0 = e$ and $[g_i] = \{g_ih \mid h \in {\rm H} \} \equiv {g_i}$H.
Iff H is normal in G, then $ghg^{-1} \in $~H so there $\exists ~h' \in {\rm H} \mid gh = h'g$ and the product and inverse of equivalent classes 
do not depend on the choice of representative for the 
class: $[g][g'] = 		
[g g']$ and $[g^{-1}] = 	
[g]^{-1}$. 
{\it Right} cosets may be defined likewise:
G$\setminus$H = $\Sigma_{g_i \not\in {\rm H}g_{i-1} }$ H$g_i$, as opposed to $g_i$H for left cosets ($\forall h \in {\rm H}$).  }.  
 Yet, without group structure no invariant notion of product exists. 
 In a minimalist way, a group is a set with (two-sided) identity and a single invertible operation, i.e., the product operation. 
 On the other hand, a ring is a set with identity and with two binary operations: addition and multiplication (the latter distributing over the former).
 Moreover, a ring is further endowed with additive monoid an abelian group already looking very much like a vector space. 
 On the other hand, within the scope of Lie theory when the quotient in a homogeneous model geometry is not by a normal subgroup
 the {\it algebraic} coset space is simply labeled a vector space as opposed to a Lie subalgebra.

	Four dimensional Eucledian space, R$^4$, represents a canonical trivialization of what a physicist refers to as a vector space.
 In special relativity, R$^4$ is equipped with some minimal extra structure to enforce causality through a 
 globally flat metric: \Mink = (R$^{1+3},~ \langle, \rangle$).   
 In general relativity, the world metric is {\it derived} from the flat metric through the introduction of vierbeins which possess one flat index and one world index.
 These are often called solder forms in analogy to the case when one interprets R$^4$ as an affine space with automorphism group GL(4,R) and ``reduces" this space to
 one with automorphism group the Lorentz group SO(3,1).  Group reduction is then embodied by the solder form which identifies the tangent space to the base manifold 
 with the vector space of special relativity thereby breaking affine and scale invariance.
 However, this is clearly a misnomer for vierbeins, as introduced by {\'E}lie Cartan, are far richer geometrical objects
 than the solder forms introduced by his pupil, Charles Ehresmman.  
 More formally, vierbeins are not exact external differentials; i.e. they are not 1-forms, but following common 
practice\cite{EguGilHan80} we will refer to them as solder ``forms" although perhaps a better descriptor would be infinite forms or simply graviforms.  

	Algebraically (in the sense of Lie algebras) and for the {\it compact} case, 
 a resolution of this issue in terms of a general infinitesimal coset space, $\cG'/\cH'$, depends on how the sub-algebra $\cH'$ is embedded in $\cG'$.  
 In the particularly tame case of a {\it reductive model geometry}, the algebra is endowed with a {\it canonical} choice of infinitesimal symmetry
 perpendicular to the stabilizer subgroup $\cH$ so that ``pure translations" (or transvections\cite{WiseD09}) may be generated by the 
 Ad($\cH$)-invariant complement of $\cH'$ in $\cG'$: $\cK' \supseteq [\cH',\cK']$, in the so-called reductive splitting of the Lie algebra.
 Such a canonical splitting leads to the notion of vector space as the algebraic quotient of $\cG'$ by $\cH'$ 
 although this quotient need not itself be a subalgebra unless the divisor subalgebra conforms an 
 ideal\footnote{\label{FN:LAvsVS} 
If an algebraic representation of the full group exists in terms of a finite set of infinitesimal generators and a {\it well behaved} exponential map,
an (infinitesimal sub-) representation of the group quotient as a vector space relaxes the requirement that it be a Lie algebra since the isotropy group 
is infinitesimally represented by a Lie sub-algebra, not necessarily an ideal in $\cG'$.  
The difference between these two may be tracked to the latter requiring ``absorption" of coset generators (see, e.g., Sharpe 2.2.7. \& 3.4.7.).
Explicitly, with $\cG' = \cH' \oplus \cJ'$:
\begin{enumerate}
	\item $[\cH',\cH'] \subset \cH'$, $\cH'$ is a sub-algebra of $\cG'$.
	\item $[\cH',\cJ'] \subset \cJ'$, a consequence of 1.
	\item $[\cJ',\cJ'] \subset \cH'$, ``absorption" of the quotient generators. 
\end{enumerate}
Reductive geometry brings about 1. and 2. while condition 3. is a special requirement leading to complete integrability (the in the sense of Frobenius theorem
which we keyed as {\it strongly involutive} condition) {\bf and} zero curvature\cite{Morita97}.
Ideal {\bf rings} in general are defined by closure and absorption; e.g., even integers under multiplication.
}
 of the original algebra, \S \ref{subs:Solder-Anatomy}.

	Geometrically (and still for the compact case), 
 this issue had led us to the rather abstract notion of a {\rm ``stratified manifold"} as a submersion of closed, totally geodesic submanifolds
 (= the fixed point set a given {\bf isometry}) generated by distinct isotropy types in the orbit space of a given 
$\cG$-space\cite{Pal&Ter88}.  
 To the best of our knowledge, a working notion of vector space to support appropriate group representations of the field content of a given theory 
 of gauge interactions with multiple stabilizer subgroups does not exist. 
 In essence, this amounts to inducing a representation of a suitable vector space from the isotropy sub-groups of the field content (see Appendix \ref{Appx:Slices}).

	Physically, 
 motivation for the above construction in flat spacetime comes from Wigner's original classification scheme for unitary irreducible representations of the 
Poincar\'e group\cite{Wigner39}. 
 On an irreducible representation of the first Casimir operator of the Poincar\'e algebra, ${\bf P}^2$, 
 representations necessarily split into massless and massive states according to $P^2 \geq 0$.  
 Choosing a non-zero momentum on the mass shell $p^2 = m^2$, the ``little" group of the double cover of the Lorentz group, SL(2,C), is the subgroup 
 which leaves $p_\mu$ invariant under $M^\dagger p_\mu M$, with $M \in$ SL (2,C). 
 For positive energy states $p_0>0$, there are {\bf two} stabilizer sub-groups:
 $\cG_{m^2>0}$ = SU(2) and $\cG_{m^2=0}$ = Spin(2) ${\supset \!\!\!\!\!\!\! \times}\, {\rm R}^2$. 
 Thus, simply labeling the irrep's by the first Casimir element, the need for stratification of the vector space to support such representations becomes manifest.
 We posit that stratification causes the breaking of the affine symmetry for spacetime bundles.
 Moreover, this is exacerbated by inclusion of the second Casimir invariant ${\bf W}^2$, with ${\bf W} \equiv *({\bf J \wedge P})$ the Pauli-Lubanski pseudo-tensor.  

	On the other hand, 
 Einstein's gravity--when viewed as a gauge theory in the language of fiber bundles--requires for the local and global actions to belong in a broad sense
 to a reductive pair of symmetry groups.
 Curved spacetime is then to be interpreted as the coset space that results from the very large, local diffeomorphism symmetry acting on the left 
 of the full bundle space quotiented by the rigid, global Lorentz symmetry acting on the right of the fibration. 
 Whereas the Lorentz group admits a double cover suitable for the representation of spinor fields as spinor bundles {\it associated} to the tangent bundle
 to spacetime, it is much less clear how such structure may be instilled upon the local symmetry of gravity: diffeomorphisms of spacetime.
 In fact, even before attempting to address the issues of causal and spin structure rather little is formally well understood of the structure of this symmetry 
 (as a group) beyond one dimension and interpretations surrogate to the notions of Lie theory for finite dimensional groups. 
 Yet, a critical view of the standard reductive G-structure framework reveals clear pitfalls when attempting to address the full geometrical structure of this 
 symmetry in 4D (see \S \ref{sec:IndReps}). 

	Broadly speaking, given a distinguished set of stabilizers for the particle content of the theory, 
 it is highly desirable to build a stratified vector space ``from the bottom up" by: 
 {\it i-} noting the existence of a {\it non-free} group action induced by multiple maximally compact stabilizer subgroups,
 {\it ii-} selecting an algebraic ideal in a suitable ring and its associated module tailored for the treatment of gravity as a ``gauge" theory 
 in the jargon of fiber bundles  
 and {\it iii-} replacing group morphisms by ring morphisms to implement the representation in a more general non-commutative algebro-geometric setting.

	This is a ambitious agenda of which we only touch ``the tip of the iceberg".  
 Tacitly assuming that the existence of distinct strata breaks the affine symmetry of spacetime,
 in this paper we build a mathematical framework to address {\it ii-} and {\it iii-} {\it via} the novel notion of a
 {\it double-sided vector space} as a conjecture for how the full bundle space of spacetime symmetries may be built. 	
 On the other hand, at this stage, we simply rely upon the notions of groupoids and stacks as a suggestion for how to address non-free group actions in the 
 ``amalgamate" of a stratified manifold (or quotient 
stack\cite{Sharpe02} to be technical) leaving a more explicit construction of the stratified double-sided vector space for future work.

 	The paper is structured as follows.  In \S \ref{sec:IndReps}, the non-linear realization of spacetime symmetries is reviewed within the framework of 
 induced representations.  This construction is then cast in the light of the diffeomorphism symmetry of the base space by inspecting 
 adjacent diffeomorphism quotients according to their filtration by degree as operators on the ideal of vanishing Dirac ``functions". 
 This sets the stage for the interpretation of the very large left symmetry of spacetime as a derivation.  
 Along the way, Cartan's construction of the frame bundle as a flat holomorph is reviewed.  
 In particular, this involves defining tautological and fundamental 1-forms while peripherally touching upon the categorification of the full construct. 
 In \S \ref{sec:Solder-Geo} the connection one-form is defined for a standard fiber bundle while focusing on the equivariant nature of the maps involved.
 \S \ref{subs:Solder-Anatomy} takes a more formal look at the algebraic construction of a holomorph and at the standard way of defining the tangent bundle 
 as a fibered product with the principal frame bundle.  Causal structure is then instilled upon the base manifold through a reductive bundle morphism.  
 This section ends with the well known algorithm for the reconstruction of the bundle space from transition maps on the base {\it via} equivariant maps.  
 \S \ref{subs:SolderReTool} constitutes the core of the paper.  
 It begins by stating the fact that the existence of a principal bundle is a necessary condition for the definition of parallel transport 
 when the topology of the fiber is non-compact.  
 A brief primer on crossed modules and their connection to not necessarily abelian group extensions and semidirect products is then given.
 As it turns out, the non-linear transformation law 
 for a section into a {\it curved spacetime bundle} naturally suggests itself for interpretation as a double-sided vector space.
 We elaborate on this idea while dwelling on whether the germ of the ``point" on the spacetime should be the vector space or the fiber.  
 While the construction of a cross-module (a.k.a. a 2-group) is convoluted because these two groups are not necessarily naturally embedded into each other;
 we conclude that the embedding for the curved spacetime are in fact physically natural and that the generator of the spacetime ``bundle" 
 is the symmetric, associative, unital enveloping algebra of the (non-compact) double cover of the Lorentz group; 
 so much for a choice of algebra generators.  
 As a byproduct of the introduction of a doubled-sided vector space, the notion of a left $\alpha$-derivation now takes center stage. 
 The deep meaning of this notion is that commutators in the Weyl algebra pick up a ring endomorphism upon commutating elements of the symmetric algebra 
 from left to right.  Furthermore, because higher Weyl algebras may be defined inductively, this construction may in principle encompass the full differential 
 structure of the solder form to arbitrary degree. 
 We close with a powerful conjecture: Summoning the left $\alpha$-derivation in the context of spacetime symmetries may constitute the key to a quantization of 
 the theory;
 so much for the poetry, let us proceed.

\section{Induced Representations of Spacetime Symmetries} 
 \label{sec:IndReps}

 	We want to cast light on the interpretation of curved spacetime as a local, {\it inhomogeneous} coset space in the so-called 
 non-linear realization of the local symmetry of gravity.  This vague statement will find its formal motivation in the main body of the paper.
 In its full glory, the non-linear realization of the full diffeomorphism group of spacetime would involve inducing a ``representation" of Diff M ($\equiv \cDM$) 
 from a maximally compact subgroup of $\cSO(1,3) (\equiv \cL$) under the implicit assumption that such representation could be defined {\it via} adjunction:
 Invoking {\it Frobenius reciprocity} so that the induced and restriction maps constitute a pair of adjoint functors. 
 Alas, the full diffeomorphism group of the fiber to spacetime is too complex to fit into a simple Lie theoretic framework (see, e.g., \cite{Morita97} pg. 234). 
 In fact, one could argue that $\cDM$ must be envisioned in an affine, coordinate-free manner\cite{Gronw97}; with the breaking of affine symmetry 
 bearing coordinates and frames\cite{Kirsch05} in the {\it linear} representation space of the tangent bundle
 {\it plus} the particle content of the theory on the base manifold. 

	A more pedestrian approach leaves out affine and $\cC^\infty$ issues by {\it linearly} representing $\cDM$ as a first order Jacobian with {\it constant}
 coefficients
\beq
 	\cDM \simeq \cDM^1 / {\cDM^2} \equiv \Aut^1 \rM, 
\label{Eq:LinearDM}
\eeq
 where the $\cDM^i$'s may be thought of as {\it distinguished} differential operators on M of degree $ \geq i$,
 and where Aut$^1$M--the space of linear functionals on M--is (tautologically) chosen to be a matrix element of Euclidean space, GL(4,R).
 Thus, finite dimensional adjacent quotients in a distinguished set of differential operators filtered by degree are generically represented by matrix elements 
 as the linear automorphism group of R$^4$.  Much of the work carried out in \S \ref{subs:SolderReTool} deals with close inspection and refinement of such morphisms
 as ring morphisms. 

 	Aside from the trivialization of such quotients, this construction is well embedded in algebraic geometry.  							
 The fundamental objects of (commutative) algebraic geometry are germs of algebraic (and in our case also differential) functions vanishing at a point $p$.
 These germs constitute the ring of {\it maximal ideals}, $m_p$, that supports the sheaf of differential functions {\it pointwise}.	
 For a finitely generated maximal ideal, the first order quotient $m_p/m_p^2$ as a quotient of  {\it exact} differentials is canonically identified with the 
 Zariski cotangent space as a vector space over $\cO_p/m_p \simeq$ R;
 where $\cO_p$ stands for the stalk of differential functions at $p$--a local ring--and 
 where taking the quotient by the ideal is commeasured with ``spectral evaluation at a point" 
 (note that when the sheaf of differential functions is replaced by the structure sheaf $\cO_{\rm Spec\;M}$, such evaluation is carried over the set of prime ideals
 instead).
 Furthermore, the quotient of adjacent powers of the ideal, $m_p^n/m_p^{n+1}$, is a finite-dimensional $\cO_p/m_p$-vector space generated by a surjective ring
 morphism from the $n^{\rm th}$ power of the symmetric algebra: Sym$^n\; m_p/m_p^2 \rightarrow m_p^n/m_p^{n+1}$ (Vakil 12.5.A).

 	Now, the key observation in relation to the linear representation of the diffeomorphism group above is that 
 this quotient is {\it formally dual} to the quotient of lowest order distinguished differential operators appearing in Eq[\ref{Eq:LinearDM}]:
\[
 	(\cDM^1 / {\cDM^2})^\vee  \equiv m_p/m_p^2. 
\] 	
 so that this ``linear representation" of $\cDM$ is just an identification with the Zariski tangent space to M. 
 Moreover, within the scope of the linear representation M equals the quotient of $\cDM$ by its ``differential little group", $\cDM^{\;1}$, 
\beq 	\rM = {\cDM / \cDM^{\;1}} ~~~~{\rm or}~~~~  \cDM \equiv  \cDM^{\;1} \sdprod \rM.
\label{eq:DiffFromSDP}
\eeq

	Let us relate this construction to Klein's original proposal.   Associated to a general 
{\it not necessarily effective} (infinitesimal) Klein pair $(\cG', \cH')$ with non-empty kernel $\cN'$ (= largest ideal of $\cG'$ contained in $\cH'$);
the ``difference of Ad($\cH$) with the identity" relates the {\it normalizer} of $\cH$ in $\cG$ to such a kernel: 
$\cN = \{ h \in \cH ~|~ {\rm Ad}(h) v - v \in \cN'$; for all $v \in \cG'$\} (Sharpe 4.3.2).  
Moreover, all the normal, closed subgroups of $\cH$ in $\cG$ may be found inductively through an algorithm that kills the original kernel 
and leaves behind the smallest ideal common to both $\cG'$ and $\cH'$ (Sharpe 4.4.1).
One then says that sequence stabilizes at the normal subgroup generated by such an ideal.

 	As it was previously remarked, the ring of differential operators of degree $ \geq i$ on M, $\cDM^{\;i}$, forms a filtration by degree:
\[
 	\cDM \supset \cDM^1 \supset \cDM^2 \supset \cDM^3 ... 
\]
 so that one may informally assign corresponding automorphism ``groups" of increasing but bounded dimension to the adjacent quotients: 
\beq
\Aut^{\;i} \rM = \cDM^{\;i}/\cDM^{\;i+1},
\label{Eq:Auts}
\eeq
 while M $\equiv$ Aut$^{\;0}$M.
 More formally, elements of the quotient $\cDM^{\;i}/\cDM^{\;i+1}$ are homogeneous differential operators of $i^{\rm th}$ degree with variable coefficients.
 When these quotients are formed by selecting a distinguished set of first order operators generated by the vector fields that form the basis of a Lie algebra 
 {\it via} their symmetric tensor product, the ensuing homogeneous monomials may be identified by degree with elements of adjacent quotients in the grading of the
 universal envelope of the Lie algebra: $(\cU\cL)^i/(\cU\cL)^{i+1}$.  Recall that the universal enveloping algebra is {\it symmetric, associative, and unital}.
 Furthermore, invoking left invariance of the basis vectors as differential operators on the group manifold yields a fixed value for the variable coefficients. 
 This is the content of the Poincar\'e-Birkhoff-Witt Theorem\cite{SternLA04}.   
 Thus, this Id assigns to the symmetrized tensor product of left invariant basis elements of a Lie algebra the role of arbitrary order differential operators on the 
 ``group manifold" {\it via} left invariance on the operators as well. 

 	Although the full diffeomorphism symmetry of the spacetime encompasses the $\cC^\infty$ structure {\it to all orders}, 
 the $\cDM^{\;n}$ quotients (as homogeneous polynomials of degree $n$) may be trivialized, order by order, 
 to arbitrarily high order by choosing a suitable matrix representations as a linear automorphism groups. 
 Accordingly, Aut$^1$M may be thought of as a {\it matrix group representation} resulting from an infinitesimal coordinate change: 
 $x^\mu \rightarrow x^\mu - \xi^\mu$
 (evidently this requires the presence of a fundamental length scale as in the absence of such there is no notion of ``infinitesimal";
 Ref. Edward Witten's remark in the introduction).
 Although Aut$^1$M is trivially identified with GL(4,R), clearly a more proper choice must enlarge the geometric
 data to enforce local causality and spin structure. 
 However, a big problem does arise if one takes the non-compact $\cL \simeq \cDM^1$ as the stabilizer subgroup for then left and right Haar measures {\it differ} 
 on the group manifold and a proper choice of automorphism group is not self-evident.
 On the other hand, since we advocate a vacuous picture of empty spacetime, it is tempting to adopt the maximally compact subgroup of $\cL$ for massive states, 
 SU(2), to induce appropriate representations of the spacetime.  Yet again, we are also strong advocates of {\it local} enforcement of causal structure.  
 We propose an potential exit to such a conundrum in \S \ref{subs:SolderReTool}.  

	Note that inducing a representation of a ``standard" vector space leads to a flat holomorph: $\cL \sdprod \Aut(\cL)$;
 the semidirect product of $\cL$ with its (inner) automorphism group Aut($\cL$). 		
 Indeed, in {\it empty} spacetimes the non-linear realization of the Poincar\'e group from its Lorentz subgroup yields an {\it isotropic embedding} of the latter 
 in the former: 
 $\cI\cS\cO(1,3) \equiv \cSO(1,3) \, {\supset \!\!\!\!\!\!\! \times} \, \cR^{1,3}$. 
 This is an extension of the Lorentz group by a vector space representation of $\cL$: 
 $\cR^{1,3} \simeq \cI\cS\cO(1,3) / \cSO(1,3)$, the {\it normal}, abelian group of (constrained) translations with Minkowski signature.
 Trivially, transition maps take values on the abelian group of translations and the corresponding equivalence class of objects invariant under such maps 
 defines the points of the manifold: Minkowski space as an abelian, homogenous coset space.
 This all makes sense since a non-linear realization of the full Diffeomorphism group should not be expected to be strongly involutive (in the sense of Frobenius):
 {\it  abelian commutators for the generators of the coset generically lead to empty, globally symmetric spacetime solutions}  
 (see remarks in Footnote \ref{FN:LAvsVS}). 
This in turn implies the non-existence\footnote{
From the perspective of algebraic geometry, the ``zero object" in $[\partial_i,\partial_j]= \emptyset$ represents {\it localization to a generic point}. 
This is inconsistent with the (strong) involutive condition on the horizontal algebra $[p,p] \subset h$ 
suggesting that $0 \subset h$ for abelian generators of motions in Minkowski space.
Even more notably, the self-consistency of non-fully symmetric spacetime solutions necessarily relax the {\bf torsion-less} condition
{\bf locally}\cite{WiseD09}.  
Indeed, it would be interesting to explore if the existence of localized torsion could restore the full integrability of the horizontal 
distribution.  Note that algebraically what we refer to full integrability is more restrictive than was is encountered in the standard
literature; e.g. the notion in Kobayashi and Nomizu corresponds to our ``weakly involutive" condition (Ref. \S \ref{subs:Solder-Anatomy}). 
} of holonomic bases for the ``world" manifold: $[\partial_\mu,\partial_\nu]= \emptyset$.  

 	Given a closed, invariant subgroup: $\cN \subset \cG$, much of the notion of homogeneous model geometry is succinctly embodied by the following 
 short exact sequence:
\[
	1 \rightarrow \cN \hookrightarrow \cG \rightharpoonup \cC \rightarrow 1,
\] 
 which not incidentally parallels the construction of a (non-necessarily principal) fiber bundle 
 $(E, F, \pi, M)_{_\cH}$     
 with structure group $\cH$ over base $M$: $\cH \hookrightarrow E \rightharpoonup M$. 
 In particular, this implies that the first map
 $\hookrightarrow$ is injective (algebraically one to one with {\it possibly} non-empty cokernel (although then $\cN$ would not be normal nor ``invariant", 
 see \S \ref{subs:SolderReTool})) while the second map $\rightharpoonup$ is surjective (algebraically onto with empty cokernel and full use of the codomain) 
 and {\it unique} although the definition is more broad in a categorical 
sense\footnote{Categorically speaking, the map 
$\hookrightarrow$ is a monomorphism while 
$\rightharpoonup$ is an epimorphism 
but these maps need not be injective nor surjective if the inclusion objects form dense sub-objects in the category. 
Moreover, by definition of a short exact sequence, the maps obey 
im [$\hookrightarrow$] = ker [$\rightharpoonup$] 
where the kernel map into the target space is not simply an object but rather a morphism of an object into the model geometry.  
Again, if the inclusion objects form dense sub-objects in the category this leads to the notion of ``bimorphism" which makes subtle contact with 
isomorphism through inclusion but not the other way around!  
Uniqueness of the epimorphism then is to be interpreted in terms of the universal property nonsense: ``unique up to unique isomorphism". 
{\bf BUT} this still depends on the particular category: in the Category of {\bf Sets} bimorphism is OK but in the category of Hausdorff topological spaces
subtle issues 
arise\cite{Freyd64}.  
We will enlarge the category to {\it locally ringed spaces} (LRS) to take full advantage of the algebro-geometric construction.
}. 

 	One of {\'E}lie Cartan's most profound insights into algebraic constructions of differential geometry 
 was to advance Klein's notion of homogeneous model space to develop the more abstract machinery of fiber bundles.  
 Yet, the colossal utility afforded by a full understanding of these notions was not readily forthcoming\cite{Sharpe97}. 
 In fact, nearly thirty years transpired before Cartan's pupil, Charles Ehresmann, partially unravelled the physical practicality of this elegant construct 
 by loosely detaching the identification of the coset space with the {\it physical spacetime} of General Relativity.  
 In this paper we are chiefly concerned with the re-identification of the coset space with the base manifold as a physical spacetime:
 {\it affine, inhomogeneous, non-compact, causal, and spin}. 

 	Today, Yang-Mills gauge theories 
 are robustly formulated within the versatile language of fiber bundles:
 Demanding Lie-group structure of the ``internal" fiber space but with tacit trivialization of the geometry of the base manifold.
 Accordingly, 	the coset space is identified {\it pointwise homeomorphically} with the linear, flat tangent space to spacetime, $\Tan M_x \simeq R^{1,3};x \in M$,
 whereas postulating the existence of a frame bundle ${\cal F} M$ supported on a fiducial vector space $V$ as an Aut($V$)-principal bundle 
 allows for an impromptu inclusion of the ``external" symmetries of General Relativity in {\it a linear} representation space.
 The tangent bundle to spacetime then is associated to ${\cal F} M$, as a pull back bundle (a fibration!)
 and this whole construction descends on spacetime proper through bundle reduction and {\it soldering} with the base manifold.
 Algebraically, the ``egg" here is {\it the germ of a point in each stratum of the spacetime} 			
 while the chicken is the frame bundle as an associated, principal Aut($V$)-bundle
 thereby identifying a choice of finite-dimensional vector space with the horizontal distribution or co-normal sheaf. 
 Differentially speaking, 
 frame bundles are ``mathematically natural" 
 and one element of the Lie group of k-jets of origin preserving 
diffeomorphisms\cite{Pal&Ter77,Eps&Thu77}. 
 As a corollary to natural bundles have finite 		%
order\cite{Pal&Ter77} the structure group of a natural bundle can always be reduced to $\cO(n)$ (in the absence of causal structure); 
 while however that of {\it smooth} fiber bundle (with the full Diff F in liu of a matrix representation) may not. 

 	The advent of the Ehresmann connection as a choice of horizontal submanifold to the entire {\it tangent space} of a principal $\cH$-bundle,
 $\cH \hookrightarrow E \rightharpoonup M$, furnishes a flexible technology generically suitable to address the internal symmetries of particle physics 
 where loosely speaking  ``spontaneous symmetry breaking" corresponds to doing physics in a coset space;
 i.e., {\it modulo} the stability group of a point in a coordinatized manifold.
 Note that such a principal $\cH$-bundle may be identified with a larger symmetry, $\cG$, with base space the coset $\cG/\cH$ as a trivial, abelian Lie group 
 and fiber space $\cH$; yet, without a globally defined $\cG$-action on the entire bundle space since an internal symmetry cannot generate the spacetime.
 Instead, a standard fiber bundle relies on the construction of the bundle space as a central group extension of the fiber 
 by the ``Lie algebra" of V as an abelian normal subgroup by demanding that all Lie brackets in the vector space generators vanish. 
 (see \S \ref{subs:Solder-Anatomy}).  This is the starting point in Ehresmann's generalization to Cartan's approach.

	Obstructions to a physically viable framework appropriate for inclusion of external symmetries lie in a 
 geometric interpretation of the notion of spontaneous symmetry breaking\cite{Trau79,Rapo&Ster84}.  
 For internal symmetries, spontaneous symmetry breaking corresponds to a map from a principal $\cG$-bundle over $M$ into a finite dimensional, {\it unstratified} 
 vector space, $V$, when the image of this map belongs to a single, ``stable" sub-orbit of $\cG$ in $V$ as the putative support of the $\cG$-representation.  
 In effect, such a map is a restriction of the orbits of $\cG$ to a sub-space of the full vector space $W \subset V$ and thus represents a {\it stratification}
 of $V$ in the formal sense of Appendix \ref{Appx:Slices}. 
 The stability group is identified with $\cH \subset \cG$ while the sub-orbit is generated by the algebraic generators of the coset $\cG'/\cH'$. 
 Such a map intertwines the right action of $\cH$ on the vertical tangent space 
 with a suitable representation of $\cH$ on a linear algebraic structure {\it locally} isomorphic to $\cH'$ as a canonical choice of ``vertical space"
or ``normal sheaf"\footnote{
\label{FN:Tor-LessMods}
In the formal sense of category theory, 
identifying the quotient with the co-normal submanifold {\it via} the co-normal sheaf is more (mathematically) ``natural" than the canonical choice of normal sheaf
as the ``internal space" generated by the stability group.  This is a non-issue for torsion-less modules (i.e., {\it `reflexive" modules} Vakil (2012) 23.2.15) 
but it does restrict the {\bf support} of the module as the set of {\bf prime ideals}.
}.  
 Iff the algebra of the larger group is identified with the full vector space, 
 the intertwining operator encodes a trivial restriction map to the normal space while the the geometric data that defines the {\it horizontal distribution} 
 as a co-normal space is non-trivially attributed to the kernel of such a map.  Furthermore, this is done {\it pointwise} throughout the spacetime.  
 It follows that if the spacetime amalgamates from distinct strata the intertwining operator is indeed a very interesting object!
 The attentive reading will recognize this simply as the definition of a connection as a ``choice" of horizontal space.
 Geometrically, spontaneous symmetry breaking is a restriction map on the full vector space while the {\it Higgs field} is given by the pullback of such an 
 intertwiner, a zero-form $\eta$, by a {\it local} section $\sigma$ of the  
principal $\cG$-bundle\cite{Trau79}: $N \stackrel{\sigma}{\rightarrow} \cG \stackrel{\eta}{\rightarrow} W$ where $N \subset M$ and $W \subset V$.
 In homological terms, this basically means that the sequence splits on the right.
 The counterpart to this construct for spacetime symmetries is given at the end of \S \ref{subs:Solder-Anatomy}.

 	The space of intertwining operators, $\cJ(V, W)$, between two {\it vector spaces} $V$ and $W$ is the fundamental object of interest in the theory of 
group representations\cite{Palais59} 
 effectively defining equivalence of vector spaces and ``naturality" of linear maps between them as {\bf equivariance} in the sense of 
 Eqs [\ref{eq:EquivI} and \ref{eq:EquivII}] below.  
 Such mathematical notion of naturality constitutes a key step in the categorification of Lie theoretic notions;
 particularly in the abstraction of algebraic structures aiming to free such structures from representation issues;
 e.g., thinking of {\it representable functors} as ``geometric spaces" 
 (although useful functors such as {\it the tangent space functor} are usually selected to be representable by construction in algebraic geometry
 through the crafty use of Yoneda's Lemma: basically objects in a category may be recovered from equivariant the maps into it 
 (up to unique isomorphism, of course; see Appendix \ref{Appx:CatNatInt}).  

	For finite dimensional Lie groups, the general framework was worked out long ago by {\'E}lie through his proposition of a flat $\cG'$-valued Cartan
connection\cite{Cartan23} with empty kernel: $\alpha$.  
 Postulating the existence of a frame bundle supported on a standard vector space, i.e., a {\it principal} bundle with structure group 
 Aut($V$) = GL($V$) $ \subset \cG$, 
 prompts the Cartan connection to split: $\alpha = \beta + \gamma$, into 
 an End($V$)-valued Ehresmann connection $\beta\!\!: {\rm T} \cG \rightarrow {\rm gl}(V)$
and 
 a V-valued {\it tautological 1-form} $\gamma\!\!: {\rm T} \cG \rightarrow V$,
 with the latter mediating an ``infra-natural" identification between the support of the representation and 
 the tangent space to the base manifold as a local homogeneous quotient:
\[
 \Tan M \simeq V \times \cG /{\rm GL(V)} \simeq V \times V; 
\]
 {\bf but not yet} with the physical spacetime of General Relativity as base manifold.  That {\bf Id} must still address two contentious issues: 
 descent (e.g., gluing of affine spaces, holonomy, torsors, zero sections, group reduction of the underlying bundle space, etc) 
 and naturalness of the Id between base and double dual vector spaces.  
 The latter issue may had confronted {\'E}lie Cartan and Alexander Grothendieck views on advancing the beauty of mathematics 
 perhaps at the cost of undermining the pursuit of truth in physics.
 Classically, Cartan's view may seem to be more physically rooted;	
 yet, the flexibility and prowess afforded by Grothendieck's program may be required for a better understanding of a complete theory of Quantum Gravity.

 	Notwithstanding the unquestionable success of YMGT in presenting a unified picture of nature away from curved spacetime, 
 at a fundamental level the standard construction represents a rather incomplete picture.
 The transformation properties of a section into the diffeomorphism bundle of spacetime symmetries, Eq [\ref{Eq:NLSigma_Transf}] below, 
 require for the global and local actions to belong to a ``reductive pair" of symmetry groups.
 On the right hand,
 Lorentz transformations of the section are rigid, spacetime independent, causal structure enforcing transformations whose algebra generates the fiber
 in the standard sense of Lie groups. 
 On the left hand, the identity component of the automorphism group of Euclidean space, GL$^+$(4,R), is simply connected and does not admit a double 
cover\cite{Gronw97} so the alleged local symmetry precludes straightforward representations of finite-dimensional spinor 
fields\cite{Neeman77,HeylAl95}. 
 The former objection clearly represents an obstruction to the standard construction of an associated frame bundle as a ``principal" bundle 
 already at the classical level whereas at the quantum level the latter objection may represent an even earlier obstruction to the assembling of the 
 tangent bundle {\it pointwise} from all of the tangent spaces at each point of the base in the presence of localized spinor and matter fields.  
 These seemingly disconnected issues are in fact intertwined and disentangling their relationship lays at the core of our enterprise.
 We will advocate a picture of quantum gravity where the notion of ``empty space" is vacuous:  
 There is no gravity (nor space or time) without stable massive and spin states; indeed, how else could one define inertial observers?
 Thus, to understand quantum gravity--at least as a ``gauge" theory--one 
 must consider stable matter and spinor fields embedded in spacetime proper through the amalgamate of a stratified manifold.

 	Moreover, a better understanding of the tangent space to physical spacetime as a representable functor is essential to incorporate the two symmetries 
 of gravity in a unified framework.
 As the attentive reader may have anticipated, we will closely scrutinize the putative ``choice" of vector space as the support of the representation of
 spacetime symmetries in \S \ref{subs:SolderReTool}.
 In this vein, an algebro-geometric construction of spacetime requires that the tangent space be a fibered product of schemes and that its representable functor
 be ``made out" of Zariski sheaves (see, e.g., Vakil 2012\cite{Vakil12}: 10.1.6).


\section{The Geometry of the Solder Form: algebraic {\it vs} differential} 
	\label{sec:Solder-Geo}

 	Given a generic fiber bundle $(E, F, \pi, M)_{_\cH}$ with the standard identifications, the kernel of the derivative map, 
\[ d\pi_p: \Tan E_p \rightarrow \Tan M_{\pi(p)},
\]
 at a point $p = (x,h)$ 
 in the fiber bundle is a vertical subspace tangent to the fiber at $p$.  The total tangent space of the bundle may thus be written as a 
 Whitney sum: 
\[
 \Tan E_p = \Tan E_p^V \oplus \Tan E_p^H \cong \ker[d\pi_p] \oplus \Tan M_{\pi(p)},
\]
 where the identification of $\Tan E_p^V$ with ker$[d\pi_p]$ is {\it canonical}
 while the identification $\Tan E_p^H \cong \Tan M_{\pi(p)}$ is {\it degenerate}.

 	A working definition of an Ehresmann connection corresponds to a choice of horizontal subspace $\Tan E_p^H$ as follows.
 First, note that for a {\it principal} $\cH$-bundle with Lie algebra $\cH'$, 
 the sub-space $\Tan E^H_p$ is invariant under the right action of an element of the $\cH$ group
 (with $R_{h'} (p) = (ph') ~\equiv (x, hh'),~~p \in E$):
\[
 R_{h*}: \Tan E^H|_p \rightarrow \Tan E^H|_{ph} \cong \Tan E^H|_p 
\] so that right translation corresponds to ``orbital motion" along the fiber only.  
 When the symmetry of the {\it fiber} is non-compact as a group, Ehresmann connections are {\it only} defined for principal bundles: the existence of 
 a right action on the full bundle space makes it possible to parallel transport vector fields and to lift horizontal curves to the bundle space 
 in a self-consistent manner\cite{Morita97}; see \S \ref{subs:SolderReTool}.
 In the case of an associated {\it non-principal} bundle, e.g., in a stratified manifold, the choice of horizontal space must be modified accordingly.

 	Next, construct the annihilator of horizontal vector fields $\Theta\!: \Tan E \rightarrow \Tan E^V$ as an $\cH'$-valued one-form while {\it identifying}
 $\Tan E^V$ with $\cH'$.  It follows that the subspace spanned by vectors $u$ such that $\Theta(u) = 0$
 defines the horizontal subspace: 
\beq
\Tan E^H = {\rm Ann} \; \Theta \equiv \{u \in \Tan E \,|\, \Theta(u) = 0 \}.
\label{eq:H-space} 
\eeq

 	Invariance of this space under the right action of $\cH$ {\it generically} imposes the requirement of ``equivariance" on $\Theta$:
\beq 
\Theta \circ R_{h*} = \Ad_{_{\cH'}} (h^{-1}) \circ \Theta.
\label{eq:EquivI} 
\eeq
 Written in composition form makes manifest the intertwining property of vector spaces that makes $\Theta$ an ``equivariant one-form"; 
 i.e., $\Theta \circ R_{h*}  \equiv R^*_h \, \Theta$. 
 More concretely, one says that $\Theta$ 
 {\it intertwines the right action of $\cH$ on the {\it vertical} tangent bundle with the Adjoint representation of $\cH$ on it's own Lie algebra.}   

	 Equivariance in this sense may be compared with the left invariance of a Maurer-Cartan 1-form valued on the Lie algebra: $L^*_h \omega = \omega$.  
 In fact, a local ``pointwise"
 pull back of $\Theta$ (from $E$) to the fiber, $\pi^{-1} x$, yields the left-invariant (fundamental) one-form {\it on $\cH'$}\cite{Morita97},  
\[
\omega = \iota_x^{*} \Theta, 
\]
 where $\iota_x:\, h \mapsto \pi^{-1} x$, 
 is a right equivariant, left invariant {\it map} into the fiber space.
 I.e., $\iota_x$ is the fiber part of a local trivialization, a {\it ``diffeomorphism"}, 
 over an open neighborhood of $x \in M:~ \iota :~ U \times \cH \rightarrow \pi^{-1} (U)$,
 consistent with right translation spanning the fiber space: $\iota_x(hg) = \iota_x(h) g$ 
 and {\it invariant} by the left composite diffeomorphism, $(\iota_x \circ L_{g})^* \Theta = \iota_x^* \Theta$, induced by a change of trivialization on overlaps.
 Such a left invariance of the vertical space ``naturally" suggests an {\it algebraic} identification of $\Tan E^V$ with $\cH'$ (a.k.a. the intertwining property).

\subsection{Anatomy of a Tautology: \\ from tautological to solder form {\it via} reduced bundle homomorphism} 
	\label{subs:Solder-Anatomy}
 
 	$\Theta$--a.k.a. the {\it fundamental 1-form} on {\it E}--is in one to one correspondence with the Ehresmann connection on a {\it principal $\cH$-bundle}.
 Notably, while $\Theta$ is valued on a Lie algebra which is isomorphic to the fiber space and is tangent to the fibers in the sense that it vanishes on 
 horizontal vector fields; in the particular case of a frame bundle the complementary notion of {\it a tautological 1-form}, $\Lambda$, 
 is a horizontal one-form that {\it vanishes} on vectors along the fiber direction,
\beq
	{\rm Ann} \; \Lambda  = \{u \in \Tan E \,|\, \Lambda(u) = 0 \} \equiv \Tan E^V  
\label{eq:V-space} 
\eeq
 (compare with Eq [\ref{eq:H-space}]), while taking values on 
the quotient\footnote{
	\label{FN:SubObj&Qs}
Such quotients tend to be better-behaved than sub-objects (note that this is a categorical distinction in the SSE)
for coherent sheaves which generalize the notion of vector bundle 
(14.1.9, 17.7, 14.5 Module-like constructions \& 3.5).
In algebraic geometry, vector bundles do not conform an abelian category so one must restrict the category to {\it (quasi)coherent sheaves}: 
the category such that maps between locally-free sheaves have well behaved cokernels as $\cO_X$-modules.}
 of a larger algebra by the algebra of the fiber: the {\it associative, unital algebra of endomorphisms} of the support of the representation, gl($V$).
 Purportedly, such a ``large" group is endowed {\it a priori} with algebraic generators for the quotient plus additional generators for the fiber
 as an Ad-invariant sub-module of the principal bundle.  
 Let us elaborate on these issues. 	

	At a fundamental level, Cartan's construction involves the existence of a principal Aut($V$)-Frame bundle. 
 By construction, such a principal bundle, $(E= \cG, F= {\rm Aut}(V), \pi, M)_{_\cH}$, is the holomorph ensuing from the {\it central}, algebraic group extension 
 of the vector space by inner automorphisms (since the extension is trivial, this is insensitive to the order of the arguments in the Ext-functor;
 i.e, ref. Eq[\ref{Eq:Struc-Morph}], below).  This, in turn, leads to the vector space playing the role of {\it ideal} for the bundle itself.
 More formally, this holomorph results from taking V to be a necessarily abelian {\it right} 
$\cL_v$-module\footnote{
\label{FN:Def-Mod}
Recall that a right $\ell$-module for an associative algebra $\cU$ is a vector space $\cM$ over a field $\psi$ together with a binary operation
$\Psi: \cM \times \cU \rightarrow \cM$ which is 
right and left distributive with respect to addition, 
commutative and associative with respect to ``scalar" product $\beta \in \psi$ 
and where a multiplicative unit exists on the right.  
}
 which is at the same time a (trivial) Lie algebra $\cL_v$
 such that the module mapping $x \rightarrow xl$ is the {\it inner derivation} in V defined by inner adjoint action (see, e.g., \cite{Jacobson62} pg. 17-18) 
 on {\it implicit} left invariant vector fields 
(following Jacobson\cite{Jacobson62}, the reader is encouraged to read the equations from right to left; such practice in ambidexterity will come in handy when 
 building hermitian operators in \S \ref{sec:TheConjecture}): 

	Define:
\bea
	(x\rD) y + x (y\rD) 	& = 	&  	(xy) \rD		\cr
	\pm xay + xya - axy 	& = 	& 	(xy) \rD_a ~~~~ ~|~ R_a - L_a \equiv \rD_a ~~{\rm :.e.g}
	\cr 
	.(x\rD_a) y + x (y\rD_a)  	& =	& 	
\eea
 Next, map: $x \mapsto xl$, under: $0 = [xy]$, while assuming: $(xl) y + x (yl) = xyl$, so that 
\bea 
 	xyl - yxl 		& = 	& 	[xy]l 	\cr
x(yl) + (xl)y - y(xl) - (yl)x	& = 	& 		\cr
	.[(xl),y] + [x,(yl)] 	& =	& 
\label{Eq:Der-via-ad}
\eea 
 For later reference, we now include a slight generalization of the ``plain vanilla" derivation above.  Simply assuming the following ``right $\alpha$-derivation" 
 rule:  $(xl) (y_\alpha) + x (yl) = xyl$, 
 which has the informal meaning that commuting the $l$ past the $y$ from right to left makes it ``pick up an $\alpha$ on the right", we get:
\bea 
	xyl - yxl 					& = 	& 	[xy] l 		\cr
x(yl) + (xl)(y_\alpha) - y(xl) - (yl)(x_\alpha)		& = 	& \cr
	.[x,(yl)]_\alpha + [(xl),y]_\alpha 		& =	& 
\label{Eq:AlpDer-via-ad}
\eea 			
 Note that commuting $x$'s and $y$'s do not pick up $\alpha$'s because $l$ is ``doing" the derivation.  This leads to the following ``right-$\alpha$ commutation rule"
\beq 	xy - yx_\alpha = [x,y]_\alpha 
\label{Eq:RightAlfComm}
\eeq

	Back to the algebraic construction of the frame bundle (and to reading Eq's from left to right!), 
 this is given by the split extension of the derivation algebra of V by V obeying the following short exact sequence:
\beq
	0 \rightarrow V  \hookrightarrow V \oplus \; \cD(V) \rightharpoonup \cD(V) \rightarrow 0,
\label{Eq:AlgHolom} 
\eeq
 In fact, V represents a {\it two sided ideal} of the Lie algebra embodied by the holomorph (essentially two-sided absorption by generators of the coset; 
 ref. footnote \ref{FN:LAvsVS}).  
 Recall that such an ideal leads to a normal subgroup {\it via} the exp$_x$ map {\it near} 0 when such a map is well defined; i.e., {\it modulo infinite 
 dimensional group structure.} 
 Furthermore, conditioning the vector space to be abelian as a sub-group restricts the extension to inner automorphisms of V, 
 $\Inn(V)$ since the derivation algebra is, by definition, the Lie algebra of linear transformations; 
 i.e., with trivial kernel as the center $\cZ$ of the full space and empty co-kernel ($\equiv {\rm Out}(\cG)$) by the Ad($s$) map:
\beq
	1 \rightarrow \cZ(\cG) \hookrightarrow V \stackrel{{\rm Ad}(s)~}{\longrightarrow} \Aut(V) \rightharpoonup \Out(V) \rightarrow 1.
\label{Eq:NonAbGrpExt} 
\eeq 
 The ambiguity between trivial Lie algebra, abelian group and vector space is prevalent in the literature.  We aim to resolve this issue algebraically, below. 
 Meanwhile, note that the algebraic statements above are robust while those referring to the group as a vector space are not.

 	Group-wise, relaxing the abelian condition and enabling the ``vector space" to be normal as a sub-group leads to non-empty outer automorphism group 
 (see below).
 Juxtaposing this with the extension of a non-trivial stabilizer sub-group, $\cH$, {\it via} adjunction, a frame bundle is a trivialized construction
 of a {\it flat} holomorph since the fundamental building block is a ``standard" vector space which does not possess a rich algebraic structure: 
 {\it it only need be additive abelian as a right $\cL_v$-module.}  

 	The above construction corresponds to a central group extension of $\cH$ = Aut(V) by $\cN$ = V: $\cG = \cN \sdprod \cH$; 
 or equivalently, to a surjective group homomorphism 
\[t: \cN \rightarrow \cH ~ | ~ \Ker(t) = \Id_\cN,
\] with trivial kernel at the center of $\cN$.
 A generalization of this construction with non-trivial kernel leads to a non-trivial co-kernel as the outer automorphism group in Eq [\ref{Eq:NonAbGrpExt}].  
 This is related to cross-modules and Lie 2-groups (\S \ref{subs:SolderReTool}).
 Yet, any notion of ``non-central extension" by a subgroup is non-existent, at least not in the language of groups. 
 In either case, the interpretation of $\cN$ as a ``vector space" runs into trouble because of it's non-abelian nature but it is this case that may be of
 interest to us in the interpretation of spacetime as a coset space.

 	The tangent space to the homogeneous model geometry is identified with an {\it associated (pullback or inverse image) vector} bundle
 (note that under the equivalence relation: $[p ,v ] \sim [ph,h^{-1}v]$ where $p\in E,~~ h \in \cH$ and $v\in \cG'/\cH'$ 
 the dimensions of the tangent bundle on the r.h.s. match twice the dimension of the (algebraic) group quotient on the l.h.s. as expected; 
 i.e., dim(E) is reduced to dim($\cH'/\cG'$) under ``$\sim$".):
\[
	\Tan [\cG/\cH] \simeq E \times_{_{\cH}} [\cG'/\cH'], 	
\]
 but this Id is {\it not} canonical; it is determined only up to the Adjoint action of $\cH$ on the algebraic coset, Ad$_{(\cG'/\cH')} (h)$, 
 as a vector space {\it not} as a Lie algebra which ensues from the {\it right invariance} of the horizontal distribution (see \cite{Sharpe97} pg. 163).   
 Thus, this tangent bundle is {\it not} a principal bundle but an associated vector bundle under a suitable equivalence relation with ``target object" 
 $\cH$ = Aut(V) in the fibered product 
sense\footnote{
\label{FN:Cats}
A more categorical, algebro-geometric construction of such a (co)tangent bundle as a {\bf fibered product} implicitly involves an underlying 
``structure morphism" from the vector space $V$, as an $\cR$-module, to an $\cR$-scheme $\cS_{_\cR}$: $V \rightarrow \cS_{_\cR}$, where $\cR$ is a field 
 or a ring or a scheme in increasing degrees of generality (see 7.3.4 of Vakil; also, 10.3.: fibers of morphisms and pulling back families). 
The fibered (pullback or inverse image) product
 then occurs over such a scheme {\it locally}:
\[
 \Tan^* M \simeq V \times_{_\cS} [\cG/{\rm GL}(V)],
\]
 and this in turn implies yet another formal map from the group quotient to the {\bf structure sheaf} in the Zariski topology:
 \[
 \cG/{\rm GL}(V)  \longrightarrow \cS.
\]
Pullback or inverse image diagram refers to the commutative diagram in 2.3.6. of Vakil illustrating the notion of universality up to ``unique isomorphism".
{\it Pullback sheaf} is reserved for {\it quasi-coherent sheaves} while inverse image sheaf is left adjoint to pushforward for general sheaves (3.6.1.).
For classical differentiable manifolds, given a continuous map of manifolds: $f:X \rightarrow Y$, one can always pullback the sheaf of (differentiable) 
functions but the notion of ``function" is more subtle in the category of schemes. 
Given a {\it morphism} of schemes as topological spaces $f:X \rightarrow Y$, 
this carries a pullback map of structure sheaves as a morphisms of {\it locally ringed spaces $f^*:\cO_Y \rightarrow \cO_X$; see also Definition 7.3.1.}  
We shall say a bit more about this structure morphism in a future publication. 
}.  
 
	Furthermore, in Cartan's picture, the tautological one form is an {\it equivariant operator} intertwining the right action of $\cH$
 on the horizontal distribution of the frame bundle with the {\it fundamental} representation of the stabilizer on the algebraic quotient:
\beq  
	\Lambda \circ R_{h*} = L_{h^{-1}} \circ \Lambda.
\label{eq:EquivII}
\eeq
 Evidently, such a decomposition assumes algebraic structure (not nessec. group) of the underlying quotient as a vector space, 
 not necessarily as a Lie sub-algebra; 
see Footnote [\ref{FN:LAvsVS}]. 
 Moreover, consistent with the definition of $\Theta$ as annihilator of horizontal vector fields, this decomposition implicitly requires {\it right invariance}
 of the vector fields that span the horizontal distribution: $\pi R_{h} = \pi$.
 Thus, we may equivalently write $\Lambda \circ R^*_h = \Ad_{h^{-1}} \circ \Lambda$.

 	One may now elegantly re-assemble fundamental and tautological one forms into a single {\it Cartan connection} {\it over the vector space} (not a group): 
\[	\Xi = \Theta + \Lambda,
\] which is manifestly equivariant (in the sense of Eq [\ref{eq:EquivI}]) under the action of $\cH$ = Aut (V) \cite{Sharpe97}.
 Note that the interpretation of $\Lambda$ here as an equivariant operator is much more subtle than before.  
 Indeed, this makes implicit use of the fact that the Cartan connection is an {\it absolute parallelism} of the frame bundle as a central group extension.

	Consider now a {\it not necessarily closed} subgroup of $\cH \supset \cL$, and build the following fibered product: 
\[
	E \times_{_{\cH}} [\cH/\cL] \longrightarrow  E/\cL.
\]
 Standard lore \cite{Kob&Nom56} posits that 
 the obstruction to a reduction of the structure group from $\cH$ to $\cL$ lies in the admission of sections into this associated bundle with putative fiber 
 the equivalent class [$\cH/\cL$] invariant under left translation by a$^{-1} \in \cH$ (corresponding to group action by the fundamental representation).  
 Furthermore, the ``canonical" Id  with E/$\cL$ through 
 the map: (u, a$\xi_0) \rightarrow$ ua $\in$ E/$\cL$, purposefully forgets the origin of the coset space $\xi_0 \in \cH/\cL$ 
 (this is an example of a ``forgetful functor" mapping elements of the associated bundle to (affine) elements of E/$\cL$).  
 Thus, in this picture {\it causal structure ensues} locally if the associated bundle with reduced Lorentz group structure, $\cL$, admits a local section
 but a ``gravitational field" is not yet present since the holomorph was {\it flat} to begin with.  
 This is the first manifestation of the {\bf dual character} of the solder form; it is an entity that lives on both, the tangent space; i.e., in the flat holomorph,
 {\it and} on the base, curved spacetime.  Clearly the latter is endowed with a much richer geometrical structure as an infinite dimensional Fock space.

	In fiber bundle language, the pullback of the tautological 1-form along an $\cL$-bundle morphism \cite{Sternb85}, 
\[ f^{w}\!\!:\, \cQ_\cL \rightarrow E_{\rm Aut(V)}, 
\]
 yields the solder form on a {\it causal, geodesic and holonomic} bundle:
\beq f^{w*} \Lambda = \Upsilon, \label{Eq:Taut2Solder} 
\eeq thereby 
 reducing the structure group of the frame bundle from Aut(V) to $\cO$(V), the orthogonal group of the putative vector space which supports the representation.

 	This so-called bundle reduction presumes that the fiber algebra of the frame bundle encompasses the algebra of the $\cQ$-fiber 
 as an Ad-$\cH$-invariant sub-module of gl($V$), so that the frame bundle is formally endowed with a reductive {\bf G}-structure.  
 Such a reductive model geometry anchors the frame bundle to the tangent bundle of the base manifold:
 {\it locally} Id'ing the {\it algebraic quotient} {\it pointwise, homeomorphically} with the support of the representation as a physical spacetime, 
 i.e., $\cD' /\cL' \simeq R^{1,3}$.  
 {\it The corresponding curtailment from tautological to solder form is the enactment of spontaneous symmetry breaking in the context of spacetime symmetries.}

	On the other hand, it is well known that one can {\it reconstruct} the bundle from transition maps as follows.
 Given a partition of unity on the base manifold as a topological space and charts subordinate to the open cover indexed by such a partition,
\beq
	\psi_i:U_i \times H \rightarrow \pi^{-1} (U_i),
\label{eq:LocTriv} \eeq
 each such diffeomorphism as {\it inclusion} into the fiber bundle must satisfy {\it right} equivariance on the fibers.
 Thus, $\psi_i (u_i, h_i) = \psi_i (u_i, e) h_i$ so that {\it a given chart corresponds to a local section} $\sigma$ over $U_i$: $\psi (u_i, h) = \sigma(u)h$.
 On overlaps, two sections from different trivializations must agree on the fiber
\[
	\psi_i \rightarrow \pi^{-1} (U_i) = \sigma_i(u_{ij}) \, h_i = \sigma_j (u_{ji}) \, h_j  = \pi^{-1} (U_j) \leftarrow \psi_j
\]
 so that $\sigma_i$ and $\sigma_j$ are related by a smooth map, k: $U_{ij} \rightarrow \cH$, acting on the {\it right} of sections: 
 $\sigma_j = \sigma_i k$ or 
\beq
k = \sigma_i^{-1} \sigma_j |_{u_{ij}}.
\label{eq:CompSec's} \eeq
 Thus, composition of sections on overlaps is in one to one correspondence with the transition functions and one can {\it reconstruct} the bundle from transition 
 maps as previously advertised.  
 The crucial key in this result is the existence of ``equivariant structure" on the underlying bundle 
(or {\it gerbe} \cite{SharpeE99,SharpeE03,SharpeE01}).  
 This is a powerful statement in the context of category theory since, according to Yoneda's Lemma, the object in the category may be recovered by knowing 
 the maps into it (up to unique isomorphism, of course).

\subsection{Anatomy of a Tautology II: a geometric re-tooling of the solder form}
 \label{subs:SolderReTool}

 	Recall that when the symmetry of the fiber in non-compact, the notion of parallel transport is only defined for principal bundles\cite{Morita97}.
 On the other hand, we concluded in the last section that the bundle may be reconstructed from transition maps which in turn are surrogate to the holonomies
 of the {\it base} manifold.  The latter must take into account manifold stratification and also include the monodromies that result from 
 ``integrating over the boundary" of non-contractible (i.e., disconnected pieces on the manifold), embedded stable massive and spin states. 
 In this paper, we address the former issue but leave the latter for a future publication.

	Assuming simply that the ``large" group $\cD$ has $\cL$ as a closed, {\it normal} subgroup, 
 the following principal fiber bundle may be readily constructed
 (with its corresponding short exact sequence):
\[
 (\cD, \cL, \pi, \cQ \equiv \cD/\cL)_{_\cL};~~~ 1 \rightarrow \cL \hookrightarrow \cD \rightharpoonup \cQ \rightarrow 1,
\]
 as a (necessarily central) {\it group} extension of the quotient $\cQ$ by $\cL$ (= Ext($\cQ, \cL$)): 
\beq	\cD = \cQ \sdprod_{\cL \rightarrow {\rm Aut} (\cQ)} \; \cL
\label{Eq:Struc-Morph}
\eeq
 where the standard group morphism implicit in the semi-direct product\cite{McLa&Birk93}: $\cL \rightarrow {\rm Aut}(\cQ)$, is exhibited 
 and furthermore particular emphasis is placed on the implicit map\cite{Weib94} 
\beq	\alpha: \cQ \rightarrow {\rm Aut}(\cL),
\label{Eq:Inv-Struc-Morph}
\eeq 
 as a {\it section} (not necessarily a group homomorphism) enforcing part of the consistency conditions for a Lie {\it 2-group} $\cG$\cite{Kapu&Thor13}.
 Notably, the short exact sequence splits only if kernel of the map 
\beq	t:~ \cL \rightarrow \cQ 
\label{Eq:t-map} 
\eeq is central
\cite{Neeb05}. 

	We can {\it inequivalently} reverse this construction leading to a more standard picture of the semidirect product as an extension of $\cL$ by an
 {\it abelian} group $\cU$ (which we coined as the ``vector space" in Eq [\ref{Eq:AlgHolom}]: an algebraic extension {\it via} {\bf a} derivation algebra):
\beq
 1 \rightarrow \cU \hookrightarrow \cL \sdprod_{\cU \rightarrow {\rm Aut} (\cL)} \; \cU \rightharpoonup \cL \rightarrow 1,
\label{Eq:Abe-Ext}
\eeq
 but, as noted, this is not equivalent to the normal, {\it non-abelian} group extension of the quotient ``group". 
 
 In fact, the fundamental difference resides in the kernel of $t$-map, Eq [\ref{Eq:t-map}].  

	Now, the solder form may be pictorially represented by its ``non-linear transformation law"\cite{Lord&Gosw88,Sard02,Kirsch05}: 
\beq 
	\cD \;\sigma\; \cL^{-1} = \sigma'
\label{Eq:NLSigma_Transf}
\eeq 	
 where $\sigma (\xi) \equiv \exp(\xi^j I_j)$ is a section of the fiber bundle $\cD$ with fiber $\cL$ over the coset $\cD/\cL$ and $\cD$ acts on $\sigma$ 
{\it locally} from the left\footnote{
We thank Juan Maldacena for this key observation on the melding of the local and global symmetry groups for gravity while highlighting the fact that because 
$\cD$ is induced from $\cL$, {\it both} of these enforce local causality: the former at the microscopic level and the latter at the macroscopic level.
} while $\cL$ acts on $\sigma$ {\it globally} from the 
right\footnote{
Note that local and global roles here are in fact exchanged from what is encountered in the standard literature on non-linear realizations of spacetime symmetries.
Given a principal H-bundle G(H,G/H) with (large) group structure (i.e., a holomorph), 
``coordinates" in the full bundle space are related to the section through right multiplication by an [arbitrary group] element of the 
fiber\cite{Lord&Gosw88}.  
A full bundle diffeomorphism is implemented through left translation by an element of the group, $g$, embodied by the bundle as a holomorph.  
Such a translation moves both the original base point and the coordinates on the fiber. 
Standard lore poses that to ``compensate
\cite{ColWesZum69,Weinberg96,Lamb&West07}" for the motion along the fiber, right multiplication on the translated section by a fiber element $h$ is required.
Thus, base point translation gets ``dumped"  into ``spacetime dependencies" for the translated fiber and coset elements as follows: 
$\sigma(\xi) \mapsto g_0 \sigma(\xi')$ and $\sigma(\xi') \mapsto \sigma'(\xi')h(\xi',g)$ where the prime refers to the translated coset representative.
Thus, identifying coset representatives $\xi$ with base point coordinates, H can be seen as the local group for its dependence on $\xi$.
On the other hand, looking at elements of $\cD'$ as the {\it differential generators of the spacetime} naturally leads to its interpretation as the (infinitesimal) 
local group whereas $\cL'$ can be seen as generating rigid global transformation that enforce (local) causality.  
This is the interpretation that we will give to the meaning of Eq[\ref{Eq:NLSigma_Transf}].
}.  
 The crucial issue here is that left action by $\cD$ does not preserve the fibers pointwise on the base; 
 instead it induces a {\it diffeomorphism of the entire bundle} and in particular a change in the algebraic coset representatives, 
 $\xi_i$, that are endowed with the transformation properties of gravitational 
fields\cite{Kirsch05}\footnote{In fact, the quotient of {\it analytic} diffeomorphisms by $\cL$ yields {\it spacetime} coordinates, 
the vierbein and an ``infinite tower of `generalized' connections" as {\it dynamical} gravitational fields in the spacetime proper\cite{Boul&Kirs06}.
This complements our analysis here on the (algebraic and non-commutative) geometry of the solder form.
}.  
 This is in essence what defines our {\it geometric re-tooling} of the solder form: a section into the very large ``diffeomorphism space" of spacetime 
 embodied by a {\bf non-central extension} of the quotient space as a quasi-group (in fact a ring), $\cQ$, by the Lorentz group $\cL$; 
c.f., Eq [\ref{Eq:Struc-Morph}].  On the other hand, we are strong advocates of interpreting the fiber as the ``egg" that generates the entire bundle space 
 in some suitable fashion.  Thus, once we setup the construct below in the fashion devised by Elie Cartan for a ``principal" frame bundle, 
 it will be reversed and the bundle space will be induced from the algebra of the fiber for comparison and to try to make ends meet.
 Recall that whereas we have a fair understanding of the structure of $\cL$ as a non-compact Lie group, the structure of the quotient $\cQ$ and of $\cD$
 as ``groups" is uncertain at best and open to debate.  This seems to preclude a well behaved embedding into a {\it standard} fiber bundle construction; 
 unless, that is, one calls upon groupoids and non-commutative algebra.

 	$\cD$irect inspection of Eq[\ref{Eq:NLSigma_Transf}] suggests that the section could be built algebraically
{\it via} non-commutative ring theory\cite{Lam90}
 as a ``double-sided vector
space"\cite{Patrick00}--or possibly as a
{\it Frobenius bimodule}\cite{Pappa03}--for which the right and left ``scalar" actions differ in the prescribed 
manner--left action augmented from that on the right by an {\it left-sided algebra automorphism}: $\alpha$--while 
 being subject to the interpretation of the coset $\cD/\cL \equiv \cQ$ as a spacetime.
 Thus, following the algorithm for the algebraic building of the frame bundle as a holomorph, Eq[\ref{Eq:AlgHolom}],
 we take $\cQ$ as a (necess. abelian) {\it right} $\cQ_v$-module which is at the same time a {\it skew (non-Lie) K-algebra} $\cQ'_v$ such that 
 the module mapping $x \mapsto \wp x$ is a {\it left $\alpha$-derivation} ($\wp$ now moving from left to right) in an {\bf associative} algebra $\cA$:
\[ 
	\lalpDer (ab) = \lalpDer (a) \; b + \alpha(a) \; \lalpDer (b),
\]
 where $a,b \in \cA$, $\alpha =$ End $\cA$, so that  
\beq
	 \lalpDer a =  \lalpDer (a) + \alpha(a) \; \lalpDer 
\label{Eq:NCDR-Alg} \eeq
 as a {\it non-central} ring generalization of a Weyl algebra$^{\dagger}$. 	
 The definition of the algebra is given by David Patrick\cite{Patrick00}:
 A skewed (or twisted) K-algebra is a ring A, together with an {\it injective ring homomorphism:} $K \rightarrow A$, where elements of K may be interpreted as 
 the scalars of A; yet, K is not constrained to be central in A so the scalars have different left and right actions on A.
 Such an algebra results from the free polynomial ring in a single indeterminate with coefficients in A, ${A \langle x \rangle}$,
 cut out by the ideal generated by the left $\alpha$-derivation\cite{Patrick00}:
\beq
	\cA_1 [x;\;\alpha, \lalpDer] = {{A \langle x \rangle}\over{  ( x a - \alpha(a) x - \lalpDer(a) ) }},  
\label{Eq:LeftOreExt}
\eeq
 where $a$ ranges over all elements of $A$.  This is a left-Ore extension in a single variable, $x$, 
 although--modulo the $\alpha$-morphism--$x$ and $\lalpDer$ are in some sense ``conjugate" to each other; see Eq[\ref{Eq:AsymConjugate}] below.
 Moreover, as any Weyl algebra does, this ``first" (twisted/skewed) K-algebra, $\cA_1$, generates higher order skewed-Weyl algebras inductively:
\beq 	\cA_n(k) = \cA_1( \cA_{n-1}(k) ).
\label{Eq:Induct-Sk-Weyls}
\eeq 
 Quite remarkably, the mere fact that $\lalpDer$ is a derivation, 
enables\cite{Jacobson62} the existence of an algebraic (not necess. flat) holomorph in the sense of Eq [\ref{Eq:AlgHolom}], above. 

 	We may now assemble the bundle algebraically   
 over the non-commutative ring {\it via} the following short exact sequence:
\beq
 	0 \rightarrow \cQ'  \hookrightarrow \cQ' \oplus \lalpDer(\cQ') \rightharpoonup \lalpDer(\cQ') \rightarrow 0.	
\label{Eq:AlgST-Quot} 
\eeq
 Yet, this obviates the fact that inductively speaking (in the sense of representation theory) in principle the egg here is a maximally compact 
 (sub-)group of $\cL$, i.e., SU(2) and not the vector space, or rather,
 the {\it non-commutative} ring module that results in the case of the algebraic spacetime embodied in Eq [\ref{Eq:AlgST-Quot}].  


	In this spirit, let us reverse this construction and start with the algebra of the fiber instead: $\cL'$. 
 Thus, we take $\cL'$ as a {\it right} $\cL_v$-module which is at the same time a {\it skew K-algebra} $\cL'_v$ such that 
 the module mapping $x \rightarrow \ell x$ is a {\it left $\alpha$-derivation} in an {\bf associative} algebra $\cA$; e.g., Eq[\ref{Eq:NCDR-Alg}].
 The ensuing algebraic sequences should be compared side by side with the group sequences for the crossed-modules embedded in the semidirect products, 
 Eqs[\ref{Eq:Struc-Morph}, \ref{Eq:Abe-Ext}],
 while re-interpreting the ``vector space" as a quotient $\cQ$ with group structure: 
\bea
	0 \rightarrow \cL'  \hookrightarrow  &  \cL' \oplus \lalpDer(\cL')  &  \rightharpoonup \lalpDer(\cL') \rightarrow 0		\label{Eq:AlgST-Fiber}  \\
&&\cr
 1 \rightarrow \cL \hookrightarrow  &  \cQ \sdprod_{\cL \rightarrow {\rm Aut} (\cQ)} \; \cL  &  \rightharpoonup \cQ \rightarrow 1 	\label{Eq:NC-Ext}  \\
&&\cr
	0 \rightarrow \cQ'  \hookrightarrow  &  \cQ' \oplus \lalpDer(\cQ')  &  \rightharpoonup \lalpDer(\cQ') \rightarrow 0  	\cr 	
&&\cr
 1 \rightarrow \cQ \hookrightarrow  &  \cL \sdprod_{\cQ \rightarrow {\rm Aut} (\cL)} \; \cQ  &  \rightharpoonup \cL \rightarrow 1 	
\nonumber
\eea
 Yet again, 
 inducing a representation from the non-compact $\cL$ brushes over the need, in the sense of Frobenius, for a maximally compact (sub-)group of $\cL$, e.g. SU(2),
 to properly induce the representation despite the fact that imposing compactness may wipe out causal structure locally (see below).
 Globally, on the other hand, a spacetime foliation {\it via} groupoids may help alleviate this issue. 

	Recall now that the subscripts in the semidirect products represent implicit {\it structure morphisms} into a base object (i.e., a ring scheme, 
 Ref. footnote [\ref{FN:Cats}]) so that an immediate Id seems trivial:
\[ 
\Aut(\cL) \simeq  \lalpDer(\cQ') ~~~\wedge~~~ \Aut(\cQ) \simeq  \lalpDer(\cL');
\] however, this is formally incorrect since 
 Eqs[\ref{Eq:AlgST-Quot} \& \ref{Eq:AlgST-Fiber}] represent extensions of algebraic rings while
 Eqs[\ref{Eq:Abe-Ext} \& \ref{Eq:NC-Ext}] represent group extensions (of a quotient and a stabilizer respectively). 
 A quick fix to such an oversight starts with replacing automorphisms of the group by endomorphisms of the algebra:
\beq 
\End(\cL') \simeq  \lalpDer(\cQ') ~~~\wedge~~~ \End(\cQ') \simeq  \lalpDer(\cL');
\label{Eq:AlgCrosMods}
\eeq however, this is again formally incorrect since {\it a priori} neither the groups nor the algebras are trivially embedded into 
each other\footnote{In naive terms, enlarging the algebraic structure from an algebra to a group requires appropriate use of a suitable exp$_x$ map {\it near} the 
 origin when such a map is well defined.  
 In principle, this yields a one to one correspondence between endomorphisms of the (not necessarily associative) algebra and automorphisms of the group.
 However, in general going from the algebra to the group is not so straightforward.
 The data attached to an algebra is closest to that of a module: both live over a ring as two sets with operations. 
 The algebra is endowed with multiplication while the module is enlarged by an additive operation.
 The data attached to a group is closest to that of a ring.
 The group is the simplest: one set with identity and a single invertible operation (combined, these two axioms allow for the existence of an inverse).
 A ring is a set with identity and with two operations: addition and multiplication (the latter distributing over the former).
 Moreover, a commutative ring is further endowed with additive monoid an abelian group.  
}.  Instead, we will try to adhere to the construction of a crossed-module to understand more precise relations.    

	With this aim in mind, recall the following three maps: 
 the implicit structure morphism in the standard definition of the semi-direct product, call it $\beta$; 
 the $\alpha$-map as a {\it section} (not necess. a group homomorphism) and the $t$-map, respectively: 
 Eq[\ref{Eq:Struc-Morph}]: $\beta: {\cL \rightarrow {\rm Aut} (\cQ)}$, 	~	
 Eq[\ref{Eq:Inv-Struc-Morph}]: $\alpha: \cQ \rightarrow {\rm Aut}(\cL)$,~
 Eq[\ref{Eq:t-map}]: $t:~ \cL \rightarrow \cQ$.
 Remarkably, the $\alpha$-map becomes a group homomorphism if one considers {\it outer} automorphisms of a normal (not necess. abelian)
group\footnote{
The standard reference for non-abelian group extensions and their higher order cohomology groups (up to H$^3$) is the book by Charles Weibel\cite{Weib94},
in particular \S 6.6; a very nice and concise review may be found at Patrick Morandi's webpage: http://sierra.nmsu.edu/morandi/notes/GroupExtensions.pdf.
A more technical yet delightful reference and one that includes infinite dimensional groups is given by Karl-Hermann Neeb\cite{Neeb05}.
}:
\[ \alpha: \cQ \rightarrow {\cOut}(\cL).
\]
 
	A key fact is that the standard ad-map is an {\it inner} derivation\cite{Jacobson62}. 
 Thus, by enlarging the algebra of the fiber with generators for the coset this derivation can be extended by {\it ``outer endomorphisms"} of the fiber,
 $\alpha_{_{\rm Out}}$, as follows:
\beq	\wp \ell - {_\alpha\ell} \wp
	= 	\OutlalpDer \ell 
	\equiv 	\OutlalpDer (\ell) + 
\alpha_{_{\cOut}} 
(\ell) \; \OutlalpDer,
\label{Eq:Out_End-Ext}
\eeq
 where $\wp \in \cQ'$ and $\ell \in \cL'$.
 Algebraically, we end up with the following short exact sequence:
\[
	0 \rightarrow \cQ'  \hookrightarrow  \cQ' \oplus \OutlalpDer(\cL')  \rightharpoonup \OutlalpDer(\cL') \rightarrow 0, 	
\]
 along with the {\it implied} group extension:
\[
 1 \rightarrow \cQ \hookrightarrow  \cL \sdprod_{\cQ \rightarrow {\rm Out} (\cL)} \; \cQ  \rightharpoonup \cL \rightarrow 1.	
\]
 However, such a construct still does not enable {\it non-central extensions of a normal subgroup} $\cL$ by the double sided quotient $\cQ$. 
 This is one reason why we need to move from crossed-modules with group structure to non-commutative rings. 
 The other reason is that invoking cross-modules is not truly necessary since for spacetime symmetries the embeddings are all in fact {\it physically} natural.

 	In the standard fiber bundle framework, the {\it a priori} detachment of the internal symmetry from the spacetime means 
 not only that the full bundle space is not endowed with group structure 
 but also that the quotient of the full bundle space by the isotropy group is, informally, only a ``shell remnant" of the full, true spacetime.
 As remarked in the introduction, such quotient is generically identified with the flat tangent space to physical spacetime.  
 The curved spacetime is in fact made of a collection of these shells {\it ad infinitum} through the inductive structure provided by the skewed-Weyl algebras:
 $\cA_n(k) = \cA_1( \cA_{n-1}(k) )$; 
 c.f., Eqs[\ref{Eq:LeftOreExt} \& \ref{Eq:Induct-Sk-Weyls}].
 This is why vierbeins are not {\it exact} differential forms: soldering with the tangent space is akin to trivializing the spacetime as locally flat.

	On the other hand, given the generators of the Lorentz algebra, $L_{ij}$,
 the germ of a {\it fermionic stratum} in the spacetime bundle is generated by two mutually commuting copies of SU(2) through the combinations:
 $\cN_i = \half (\cJ_i + i\cK_i)$ and $\cM_i = \half (\cJ_i - i\cK_i)$ where $\cJ_i = \half \epsilon_{ijk}L_{jk}$ and $K_i = L_{0i}$
 and where these two algebras are related by two natural automorphisms: complex conjugation and spatial inversion.  
 Recall now that in algebraic geometry the image of a section belonging to the sheaf of algebraic generators at the stalk of a ``point" is the set of germs
 over some {\it open} neighborhood of the identity. 
 In a physical spacetime, such germs must enforce causal structure and encompass stable massive and spin states.

	Our proposal leads to an algorithm on how to build a non-commutative algebro-geometric version of the ``bundle space" of spacetime symmetries
 from the causal generators of the Lorentz group.
 With $\eta_{\alpha\beta}$ the Minkowski metric, such generators are defined by their Lie algebra without allusion to a particular representation:
\[
	[L_{\mu\nu}, L_{\alpha\beta}] = i \{ 	\eta_{\mu\alpha} L_{\nu\beta} + \eta_{\nu\beta} L_{\mu\alpha}
					- 	\eta_{\mu\beta} L_{\nu\alpha} - \eta_{\nu\alpha}  L_{\mu\beta} \}.
\]
 This means in particular that all that is truly meaningful in the r.h.s. of above equation is the combinatorial content of the index exchange
 (moving from left to right or viceversa).  
\begin{itemize}
\item
	Choose an appropriate ground ring where the (commutative) coefficients of the polynomial rings will take values on (may choose either R or C).
\item
 	Assign to the (non-commutative) basis elements of the Lorentz algebra the role of {\it first order} differential operators on the ``group manifold"
	by demanding {\it left invariance} of these elements as differential operators.
\item 
	Utilize the ground ring and the above operators to generate a free polynomial ring naturally filtered by degree in the non-commutative indeterminates.
\item 
	Cut out the free ring by the ``ideal" generated by the left $\alpha$-derivation, Eq[\ref{Eq:NCDR-Alg}], to provide the symmetric algebra embodied in
	Eq[\ref{Eq:LeftOreExt}].
\item 
	Utilize the ``total ordering" of the operators that the filtration endows to generate a grading of the differential quotients: $\cDM^{\;n}/\cDM^{\;n-1}$.
\item 
	Invoke the Poincar\'e-Birkhoff-Witt Theorem to identify these quotients with the {\it commutative grading} of universal envelope of the Lorentz algebra 
 	as homogenous monomials of $n^{\rm th}$ degree in the left-invariant differential operators. 
\item 
	Retool the notion of ``group manifold" as an algebro-geometric {\it scheme}. 
	Points on a group manifold are tacitly and commonly identified with the group elements as (not necess. differential) operators and 
viceversa\cite{Lord&Gosw88}. 
 	One may informally choose to view the operators in the grading of the universal envelope of the symmetric algebra 
	as a differential notion of a group manifold but not in the formal sense of Lie groups since there is no well defined notion of exp-map.
	In our modern version of the manifold, physical spacetime is formally a scheme: 
	a space whose stalks are {\it local rings}; i.e., a ``locally ringed space".
 	Technically and perhaps more faithfully, the {\it full bundle space of physical spacetime} is the free, rank 4 K-bimodule realizing 
	a non-commutative symmetric algebra as a skewed K-algebra;
	{\it a very rich and beautiful geometrical object indeed!}
	Equivalently, physical spacetime is the symmetric polynomial ring in {\it skewed differential operators} generated by the Lorentz algebra
	with constant coefficients in the ground ring (since left invariance was invoked earlier on).
  	This is a trivial statement on the particle content of the spacetime as a fermionic stratum generated by $\cJ_\mu$ and $\cK_\mu$ as prescribed above. 
\end{itemize}

 	The algebra automorphisms, $\alpha$, in Eq[\ref{Eq:NCDR-Alg}] are related to the  surjective ring endomorphisms that the skewed K-algebra
 encodes Eq[\ref{Eq:LeftOreExt}] through algorithms devised by David Patrick for non-commutative symmetric algebras\cite{Patrick00}.
 Such algorithms involve relating right ``multiplication" by elements of K, to left multiplication {\it via} a matrix representation with entries in 
 fancy algebro-geometric objects such as $(a,a)$-derivations and perfect squares where $a \in $ Aut(K).  
 Further analysis of the structure of these injective ring homomorphisms is beyond the scope of this paper.
 However, note that this procedure endows the ``higher automorphism groups" of Eq [\ref{Eq:Auts}] with far richer geometrical structure
 as elements in the grading of the symmetric algebra generated by the left $\alpha$-derivation.
 Investigation of the structure of such geometrical objects appears as a fertile ground for future research as well.

\section{A Conjecture on Quantization: baby steps in quantum ring theory}
 \label{sec:TheConjecture}

	On the basis of the given arguments, we now fancy the following conjecture:  Quantization of the curved spacetime involves a slight generalization of
 the canonical commutator brackets in flat spacetime where elements of the free polynomial ring in the Lorentz algebra generators as non-commuting indeterminates
{\it undergo an $\alpha$-mutation upon commuting elements from left to right.}	

	The deep meaning of the conjecture is that the flat spacetime symbols $x$ \& $p$ in the canonical commutator relation:
\[	[p,x] = -i\hbar,
\]
 must be replaced by the algebro-geometric objects $\wp$ \& $\ell$ as elements of the free polynomial ring generated by the Lorentz algebra.  
 When one of these objects, say $\ell$, is simply considered as an (asymmetric) inner derivation; e.g., following the procedure in 
 Eqs[\ref{Eq:Der-via-ad} \& \ref{Eq:AlpDer-via-ad}] but from left to right, 
 we obtain a left $\alpha$-skewed version of the canonical commutation relations:
\[ 	
	\wp \ell - {_\alpha\ell} \wp_{\hbar\alpha} \equiv {_\alpha[} \wp , \ell ] = - i\hbar,
\]
 which has the informal interpretation that the $\ell$ element undergoes an $\alpha$-mutation upon commuting $\wp$ through $\ell$ from left to right.
 The $\alpha$-mutation is in fact a ring endomorphism and a suitable framework to develop this notions further in that of a {\it quantum ring theory}.  

	As a quick check for consistency, recall that 
\beq
	[p,x]^\dagger = [x,p] = - [p,x]
\label{Eq:AntiHerBracket}
\eeq
and look at the anti-hermicity relation for the skewed bracket: 
\bea
	\left( {1\over{i}} ~ {_\alpha[} \wp , \ell ]_e \right)^\dagger 		&=& - \hbar	\cr
&&\cr
	i\{ (\wp \ell)^\dagger - ({_\alpha\ell} \wp)^\dagger \}			&=& 		\cr
&&\cr
	i\{ \ell^\dagger\wp^\dagger -  \wp^\dagger ({_\alpha\ell})^\dagger \} 	&=&  		\cr	
&&\cr
	i\{ \ell\wp -  \wp \ell_\alpha \} 					&=& i [ \ell , \wp ]_\alpha.
\eea
 Thus, the skewed counterpart to Eq[\ref{Eq:AntiHerBracket}] is
\beq
	{_\alpha[} \wp , \ell ]^\dagger  =  [ \ell , \wp ]_\alpha.
\label{Eq:AsymConjugate}
\eeq 
 Evidently ${_\alpha\ell} = \ell_\alpha = $ End$(\ell)$, but switching $\alpha$ from left to right 
 upon taking the hermitian operation in the context of the above relation is a lot more meaningful.
 This has the elegant interpretation that the operators $\wp$ and $\ell$ are hermitian but the skewed derivation switches ``preference" 
 from left to right $\alpha$-derivation upon the hermitian operation. 
 Since the construction of the double-sided vector space in insensitive to which action is right or left as long as the ``large" differential action is derived
 from the rigid one, this scheme is seen to be self-consistent! 
 Furthermore, 
 with due care in ambidexterity the construction of the fundamental building blocks of quantum field theory; 
 e.g., simple harmonic oscillators, are unaffected by the replacement of 
 Eq[\ref{Eq:AntiHerBracket}] by Eq[\ref{Eq:AsymConjugate}].
 
	The immediate question that arises is:  what does this mean for the rest of quantum mechanics as formulated in flat space?
 Let's take a quick look at the second order fundamental operators in the elements of the free polynomial ring,
 angular momentum and Lorentz boost operators: 
\[ L_{\mu\nu} = -i (\ell_\mu \wp_\nu - \wp_\mu \ell_\nu ).  
\]
 These are unaffected by the skewed derivation since they involve different index combinations. 
 Moreover, as is well known, their commutator algebras involve the flat metric and these too are unaffected by the skew Weyl algebra.  
 This suggest that these operators constitute a good set of generators for the rest of the spacetime via the free polynomial ring that is built from them. 
 So much for the digression. 

\section{Outlook}

	We have proposed a reformulation of the standard fiber bundle approach to address spacetime symmetries where the bundle space plays an active role 
 in encoding the geometric data.  A novel notion going by the name of a double-sided vector space proves to be a handy tool pulled out of 
 {\it non-commutative algebraic geometry} to represent sections into spacetime fiber bundles.
 In the mathematical literature, this is  motivated by the study of the non-commutative analogs of symmetric 
algebras\cite{Patrick00} 
 and their relationship to noncommutative ruled 
surfaces\cite{VDBerg96}. 		
 This is highly technical field yet unexplored by the physical community.
 In particular, we have completely omitted allusion to the issue of admissibility of symmetric algebras for a given two-sided vector 
space\cite{Patrick00}.  Yet, the conclusions of this note are independent of such admission.  



{\it Acknowledgments:}\\ 						
 	I am greatly indebted to Ryan Thorngren and Tudor Dimofte for extremely valuable and timely feedback;
 and to Professors Matilde Marcolli, Stanley Deser, Carlos Olmos, John Schwarz, Joseph Varilly, Shlomo Sternberg and Eric Verlinde for lending a kind, patient ear to my ramblings.
 I thank Sergei Gukov for suggesting that to address quantum gravity something really different should be brought about.
 I also thank Juan Maldacena for his insight into the local construction and for his hospitality while I visited the Institute for Advanced Study. 
 Lastly, I am greatly indebted with Caltech's Theory Group and their wonderful Friday Theory Seminar Series which represent a theorist many happy hours 
 of delightful intellectual feast.

\pagebreak 

\appendix


\section{Category theory, naturality and intertwining operators}
 \label{Appx:CatNatInt} 

	A (small) {\it category} is a pair of related sets: (objects, morphisms) where the morphims act on the objects.  
 A large category eases a requirement on the morphisms: these need only constitute a class, not necessarily a set.  
 A {\it functor} between two categories, with source A and target B, is the following data. 
 A map
\[ 	{\rm F: obj(A)} \rightarrow {\rm obj(B)}
\] and related morphisms in B, 
\[ 	{\rm F(m): F}(a_i) \rightarrow {\rm F(}a_j),
\] induced by arbitrary morphisms in A, 
\[ {	\rm m: a}_i \rightarrow  {\rm a}_j.
\] 
 Composition of morphisms is transitive, 
\[	{\rm for~ l:~ x} \rightarrow {\rm y ~~and~~ m: y} \rightarrow {\rm z,~~~ m} \circ l: {\rm x} \rightarrow {\rm z}
\]
 and associative: 
\[ 	{\rm (l \circ m) \circ n = l \circ (m \circ n).  }
\]
 Most importantly, for each object in the category, a$_i$, there exist an identity morphism 
\[	{\rm id_{a_i}: a_i \rightarrow a_i.}
\]

	{\bf Naturality} of a ``transformation" $\eta$: F $\rightarrow$ G between a pair of functors F,G: obj(A) $\rightarrow$ obj(B) is stated in terms of 
 the commutativity of functor induced morphisms F(m) and G(m) with ``intertwined" morphisms (transformations) of functorial objects    
 $\eta_{a_k}: F(a_k) \rightarrow G(a_k)$ on the target space: 
\[
 \eta_{a_j} \circ F(m) = G(m) \circ \eta_{a_i}.
\]
 This may be conveniently stated through the commutativity of the diagram implied by this algebraic relation. 

 	In particular, when F and G belong to the category of finite dimensional (smooth, topological or algebraic) vector spaces, natural transformations 
 are interpreted as the intertwining operators between the group (groupoid) representations embodied by F and G as (affine) vector spaces, or torsors. 
 Within a Lie theoretic framework, the category of (finite) vector spaces is interpreted as a category with a single object, 
 with invertible morphisms as group elements and with Lie algebra elements as generators of the vector space.

	Note that this is reminiscent of the equivalence relation for \v{C}ech 1-cocycle defined for an open cover $\cU = \{U_\alpha\}$: 
 g,h $\in \cG (U_{\alpha\beta})$ are equivalent iff $\exists \, f \in \cG (U_\alpha)$ such that 
\[ 	f_\alpha g_{\alpha\beta} = h_{\alpha\beta} f_\beta.
\]
 The set of equivalent classes yields the first cohomology class on the open cover with coefficients in $\cG$, $\Sigma_i [g_i] = H^1(\cU, \cG)$, 
 and the colimit over the refinement of covers turns the first \v{C}ech cohomology 
class\cite{Moerd02} over the base manifold: 
 \v{H}$^1(X,\cG) = \lim_{\rightarrow \cU} H^1(\cU,\cG)$.
 For completeness, we quote here the very important theorem from Moerdijk notes on Stacks and Gerbes (2002: Theorem 1.3):

{\noindent  {\it There is a bijective correspondence between isomorphism classes of $\cG$-torsors and cohomology classes in \v{H}$^1(X,\cG)$.}}

	So much for the digression.

	Given objects A and B in a category $\cC$ as above, {\bf Yoneda's Lemma} is the statement that a new, fully faithful functor can be built from $\cC$ to its 
 ``functor category" where objects are contravariant functors $\cC \rightarrow Sets$ and the morphisms are natural transformations of such functors (Vakil2012).
 More explicitly, 
 given a third object in the same category $C \in \cC$, this induces a set of morphisms from $C$ into $A$: Mor$(C,A)$, 
 so that given a single morphism, $f: B \rightarrow C$, from a different object $B$ into $C$ a map of sets in induced:
\[ {\rm Mor} (C,A) \rightarrow {\rm Mor} (B,A)
\]
 by composition 
\[	B \stackrel{f}{\rightarrow} C \rightarrow A
\]
 where the second map is given.
 Now, given two distinct objects $A, A' \in \cC$ and another set of morphisms
\[ j_C: {\rm Mor} (C,A) \rightarrow {\rm Mor} (C,A')
\]
 that commute with the first map of sets, the $j_C$'s are induced from a unique morphism $g:A \rightarrow A'$ for all the $C \in \cC$! 
 Furthermore, if all the $j_C$'s are bijections, $g$ is necessarily an isomorphism.
 Lastly, there is a functor $h^A: \cC \rightarrow Sets$ 
 sending $B \in \cC$ to Mor$(A,B)$ and 
 $f:B_1 \rightarrow B_2$ to ${\rm Mor} (A,B_1) \rightarrow {\rm Mor} (A,B_2)$.
 This is given by 
\[ 	[g:A \rightarrow B_1] \mapsto [f \circ g: A \rightarrow B_1 \rightarrow B_2].
\]
 Yoneda's Lemma is the statement that there is a bijection between the natural transformations $h^A \rightarrow h^B$ of covariant functors $\cC \rightarrow Sets$ 
 and the morphisms $B \rightarrow A$.


\pagebreak


\section{Left and right actions in representation theory}
 \label{Appx:LR-Act-in-RT} 

 	By definition, a (finite) representation of a (finite) group $\cH$ on a (finite) vector space $V$ is a 
homomorphism\footnote{
A homomorphism is just a $\cC^\infty$ map that respects the composition law: $\rho(gh) = \rho(g) \cdot \rho(h)$
plus possibly some algebraic or geometric structure.
} 
of $\cH$ to the group of automorphisms of $V$, $\rho\!: \cH \rightarrow \Aut(V)$.  The assignment of a continuous map from a {\it simply connected} Lie group
$\rho\!: \cH \rightarrow \cI$, i.e. a {\it Lie group homomorphism}, 
 is uniquely determined by its differential at the 
identity element\cite{Fult&Harr91}:
\[
d\rho_e\!: \Tan_e \cH \rightarrow \Tan_e \cI 
\] 		
 More 
formally, one says that any smooth map from $\cH$ is uniquely determined by its 
germ\footnote{
If $U \subset \cH$ is any {\it open} neighborhood of $e$, then $U$ generates $\cH$.
The differential structure of a variety may be readily generalized by introducing the notions of sheafs and stalks.  Whereas group representations 
yield ``the group element at the translated point" (e.g. the value of a continuous function at an adjacent point) in terms the differential map, 
by contrast the image of a section belonging to a sheaf $\cF$ at the stalk $\cF_p$ are sets of germs over some open set containing $p$ and $e$. 
The value of a section at a ``point" is thus undefined.
} at $e$.

 	As a homomorphism of Lie groups, 
 left conjugation, $\int_g (h) = ghg^{-1}$ (with $g,h\in\cH$), preserves the action of the group $\cH$ on itself by inner automorphism.
 The differential of this map yields automorphisms of the tangent space (expanding $h$ as a 1-parameter subgroup $h = h(t) \simeq e + At)$:
\[ 	\Ad_g \equiv d (\int_g)_e: \cH \rightarrow \Aut (\Tan_e \cH);~~~g,e\in\cH. 
\]
\[	d(\int_g): A \longmapsto g A g^{-1}; ~~ g \in \cH, ~~A \in \cH'.
\]
 As a map from $\cH$ to $\Aut (\Tan_e \cH)$ this is by definition a representation of $\cH$ in its own tangent space; 
 it constitutes a bookkeeping devise to track the ``difference" between left and right translations by elements of $\cH$ on elements of $\cH'$.
 In fact, the composition law for homomorphisms as $\cC^\infty$ maps: $\rho(gh) = \rho(g) \cdot \rho(h)$, also permits either left or right 
 translation as group operation in lieu of conjugation; these maps {\it respect} the action of the group $\cH$ on itself by diffeomorphism. 
 Nevertheless, automorphisms are favored over diffeomorphisms as differential structure preserving morphisms because these {\it enforce} 
 the presence of a fixed point, namely the choice of origin, ``$e$", to contain the representation locally.
 One then exploits the continuous structure of the group at the identity to build the representation. 
 The main obstruction to a locally diffeomorphic construction is the absence of {\it generic} fixed points.

	It should be emphasized that the Adjoint representation is a mixing bilinear gadget adjoining elements of the group with those on its algebra. 
 Since in order to exploit the continuous structure of the group at the identity one desires a condition purely on the differential of the homomorphism map, 
 d$\rho_e$, it is necessary go one step further and take the differential of the Adjoint map, d Ad $\equiv$ ad, to get the adjoint representation acting on
 elements of the Lie algebra alone.  
 In the context of {\it finite dimensional} Lie Groups where there is a well defined exponential map relating elements of the algebra to elements of 
 the group the procedure is a straightforward; but the absence of such a map in the infinite dimensional case, i.e. for the diffeomorphism group, 
 requires a more sophisticated (and uncertain) procedure. 

 	For finite dimensional groups, the map ad is realized as map from the tangent space into endomorphisms of the same:
\[ \ad = d \Ad_\bullet  : (\Tan_e \cH) \rightarrow \End(\Tan_e \cH). 
\]
\[
	\ad (A)  = \lim_{t \rightarrow 0} ~ d (\Ad_{g(t)}) = [A,\bullet]
\]
 where $g(t) = e^{tA} $ is a one parameter subgroup starting at the identity $e$.
 One then has $d\rho_e({\rm ad}(x)(y)) = {\rm ad}(d\rho_e(x)\rho_e(y)$. 
 Note that in both cases, the tangent spaces must be interpreted as vector spaces and not as Lie algebras {\bf per se.}  Also,
 automorphisms constitute a dense open subset of endomorphisms since the latter allow for non-invertible maps into the target space.
 Diffeomorphisms, on the other hand, add the extra requirement of differentiability of both the map and its inverse.   
 This is a stronger requirement than continuity which yields homeomorphisms of manifolds.
 For further explanation including the implicit commutative diagrams involved in these ``natural" maps,  
 the reader in referred to the beautiful exposition by Fulton and Harris\cite{Fult&Harr91}.


\section{Strata, Slices and Associated Bundles} 
 \label{Appx:Slices} 

	We proceed to the formal definition of a stratified manifold following the expositions by Borel \etal\cite{Borel60} and Palais and Terng\cite{Pal&Ter88}

	Geometrically, the issue of vector spaces as quotients by not necessarily normal subgroups leads to the following rather formal construction.
 Consider a $\cG$-manifold, M, where $\cG$ acts on M from the {\bf right}.  
 The {\it orbit space} is defined by the coset $\tilde{\rm M} \equiv {\rm M}/\cG$ and this orbit map is a {\it smooth} fiber bundle: M$/\cG \times \cG$.   
 Given a {\it closed}, not invariant/normal sub-group $\cH \subset \cG$, 
 one defines an {\bf $\cH$-slice}, $\cS$, in M as an $\cH$-invariant subset of M 
 such that, given a local section in the coset $\sigma: \cG/\cH \longrightarrow \cG \mid  \sigma(\cH) = e|_{_\cG}$,
 the {\bf map} $\cJ: \cU_{_\cH} \times \cS \rightarrow$ M given by $\cJ(u,s) = \sigma(u) \cdot s$ is a {\bf local homeomorphism} of 
 $\cU_{_\cH} \times \cS$ into an {\bf open set} in M (note that in this map $\cG$ is {\bf left} acting). 
Here $\cU_{_\cH}$ is a open neighborhood of $\cH$ in $\cG/\cH$ (recall that such a neighborhood generates the representation {\it via} 
 its germ  at the identity of $\cG/\cH$; Appendix \S \ref{Appx:LR-Act-in-RT});
 i.e., [g] = g$\cH$ is ``close" to e$\cH$ so that points in the coset $\cG/\cH$ are it {\it one to one correspondence} with points of the slice $\cS$ in M.

  	A slice at a point $x \in$ M is implicitly understood to be a $\cG_x$-slice where $\cG_x$ is the isotropy or stabilizer subgroup of $x$.
 If such a slice exists, then there is an open neighborhood of $x$ where the isotropy groups of each point are conjugate to that of $x$: $\cG_y = g\cG_x g^{-1}$ 
 for some $g \in \cG$ and $y \in \cU_x$.  This statement makes use of the local section as: $G_{\sigma(u)s} = \sigma(u)G_s\sigma(u)^{-1}$.
 Note that an $\cH$-slice (not to be confused with an $\cS$-representation)
 is the equivalent of a {\it co-normal} neighborhood of $\cH$ in $\cG$ passing through $e$: 
 transverse to $\cH$ and of dimension complementary to $\cH$ in $\cG$\cite{Frankel97}.  Thus, {\it the action is neither transitive nor free.}
 We choose to act on the right of M in order to make a distinction explicit in the construction of the associated bundle below.

	Next, recall the definition of the normalizer of $\cH$ in $\cG$ as the maximal subgroup of $\cG$ where $\cH$ is normal.
 For finite dimensional Lie groups with a well defined exponential map, such group can always be found (Ref. footnote
[\ref{FN:LAvsVS}]).
 A key fact(\cite{Borel60} pg. 157) is that the {\bf fixed point set} of a closed subgroup $\cH$ in the $\cH$-invariant subspace of orbits of type $\cG/\cH$, 
 M$_{(\cH)}$, is an $\cN_\cH$-slice and under the action of $\cN_\cH$ it is a principal 
$\cN_\cH/\cH$-bundle 
 over the restricted orbit space M/$\cH$.
 Further,
 taken as a space with right translation as group action, $\cG/\cH$, is a principal   
 $\cN_\cH/\cH$-bundle over $\cG/\cN_\cH$.
 Furthermore, the action of $\cG$ commutes with that of $\cN_\cH/\cH$ so these may be combined into a single $\cG \times \cN_\cH/\cH$-action for the  
 $\cG/\cH$-space.






\pagebreak 


{\footnotesize 

\begin{thebibliography}{99}
  
\bibitem{Klein1872} 
 	Klein, F., ``A Comparative Review of Recent Researches in Geometry", http://axXiv.org/pdf/0807.3161v1.pdf

\bibitem{EguGilHan80}  
	Eguchi, T., Gilkey, P., \& Hanson, A. ``Gravitation, Gauge Theories and Differential Geomatry", Phys. Rep. 66, \#6, (1980) 213-393 

\bibitem{WiseD09}
	Wise, D., ``Symmetric Space Cartan Connections and Gravity in Three and Four Dimensions", Sym. Int. \& Geo Met \& App (SIGMA) 5, 80 (2009) 

\bibitem{Stern&Ungar79}
	Sternberg, S. \& Ungar, T., ``Classical and pre-quantized mechanics without Lagrangians or Hamiltonians", Hadronic Journal, 1, 33 (1978).

\bibitem{SternLA04} 
	Sternberg, S. ``Lie Algebras", (2004) PDF found at:http://www.math.harvard.edu/~shlomo/

\bibitem{Gero68} 
	Gerosch, R., ``Spinor Structure of Space-Times in General Relativity. I.",  J. of Math. Phys. 9, 1739 (1968).

\bibitem{Gero70} 
	Gerosch, R., ``Domains of Dependence", J. of Math. Phys. 11, 437 (1970).

\bibitem{Pal&Ter88}
	Palais, R. \& Terng, C., ``Critical Point Theory and Submanifold Geometry", Lecture Notes in Mathematics 1353, Springer-Verlag (1988).

\bibitem{Pal&Ter77}
	Palais, R. \& Terng, C., ``Natural Bundles have Finite Order", Topology 16, 271 (1977).

\bibitem{Wigner39} 
	Wigner, E. P. (1939), ``On unitary representations of the inhomogeneous Lorentz group", Annals of Mathematics 40 (1): 149–204.

\bibitem{Gronw97}
	Gronwald, F., "Metric-Affine Gauge Theory of Gravity I: Fundamental Structure and Field Equations", 
	Int. J., Mod., Phys., D 6, 263, (1997) (axXiv:gr-qc//9702034) 

\bibitem{Kirsch05}
	Kirsch, I., "A Higgs Mechanism for Gravity", Phys. Rev. D72 (2005) 024001 (arXiv:hep-th/0503024)

\bibitem{Boul&Kirs06}
	Boulanger, N \& Kirsch, I., ``A Higgs Mechanism for Gravity. Part II: Higher Spin Connections", Phys. Rev. D73 (2006) 124023 (arXiv:hep-th/0602225)

\bibitem{ColWesZum69}
	Colleman, S., Wess, J. \& Zumino, B. ``Structure of Phenomenological Lagrangians. I."  Phys. Rev. 177, \#5, 2239.

\bibitem{Weinberg96}
	Weinberg, S. ``Quantum Theory of Fields V2: Modern Applications", Cambridge University Press, 1995. 

\bibitem{Lamb&West07} 
	Lambert, N. \& West, P. ``Duality Groups, Automorphic Forms and Higher Derivative Corrections Phys.Rev.D 75:066002 (2007) 	

\bibitem{Frankel97}
	Frankel, T., ``The Geometry of Physics: An Introduction", Cambridge University Press (1997).

\bibitem{Fati&Fran03}
	Fatibene, L. \& Francaviglia, M., ``Natural and Gauge Natural Formalism for Classical Field Theories:
	 a geometric perspective incliding spinors and gauge theories", Kluver Academic Publishers, Dordrecht/Boston/London (2003).

\bibitem{Sharpe97} 					
	Sharpe, R.W., ``Differential Geometry: Cartan's Generalization of Klein's Erlangen Program". 
	Graduated Texts in Mathematics \#166,  Springer-Verlag (1997). 

\bibitem{Sternb85} 			
	Sternberg, S., ``On the Interaction of Spin and Torsion II. The Principle of General Covariance", Ann. of Phys. 162, 85 (1985).

\bibitem{Borel60}
	Borel, A., Bredon, G., Floyd, E., Montgomery, D. \& Palais, R., ``Seminar on Transformation Groups", Annals of Mathematical Studies 46, 
 	Princeton University Press (1960). 

\bibitem{Eps&Thu77}
	Epstein, D. \& Thurston, W. (1979), ``Transformation Groups and Natural Bundles" Proc. London Math. Soc.(3) 38, 219.

\bibitem{Rapo&Ster84} 
	Rapoport, D., \& Sternberg, S., ``On the interaction of Spin and Torsion" Ann. of Phys. 158, 447 (1984).

\bibitem{HeylAl95}
	Heyl, F.W., McCrea, J.D., Mielke, E.W., Ne'eman, Y. ``Metric-affine gauge theory of gravity:
		 field equations, Noether indentities, world spinors, and the braking of dilation invariance" Phys. Rep. 258 1-171 (1995)

\bibitem{Neeman77}
	Ne'eman, Y. ``Gravitational interaction of hadrons: Band-spinor representations of GL(n,R)" Proc. Natl. Acad.Sci. USA V74, No.10, 4157-4159 (1977)

\bibitem{Trau79}
	Trautman, A., ``The Geometry of Gauge Fields" Czech. J. of Phys. 29, 107 (1979).

\bibitem{Ehre50} 
	Ehresmann, C., ``Les connexions infinit\'{e}simales dans un space fibr\'{e} diff\'{e}rentialble", Colloque de Topologie, Bruxelles (1950) pag 29-55

\bibitem{Cartan23} 
	Cartan, E., ``Sur les vari\'{e}t\'{e}s \`{a} connexion affine et la th\'{e}orie de la relativit\'{e} g\'{e}n\'{e}ralis\'{e}e",
	ann. Ec. Norm. 40 (1923), 325-412; ibid. 41 (1924), 1-25; ibid. 42 (1925), 205-241

\bibitem{Jacobson62}
	Jacobson, N., ``Lie Algebras", Interscience Tracts in Pure and Applied Mathematics No. 10, John Wiley \& Sons, New York $\bullet$ London (1962).

\bibitem{Lam90}
	Lam, T.Y., ``A First Course in Noncommutative Rings"  Graduate Texts in Mathematics \# 131, Springer (1990).

\bibitem{Patrick00}
	Patrick, D., ``Noncommutative Symmetric Algebras of Two-Sided Vector Spaces" J. of Alg. 233, 16-36 (2000).

\bibitem{Pappa03}
	Pappacena, C.J., ``Frobenius bimodules between noncommutative spaces" J. of Alg. 275, 675-731 (2004).

\bibitem{Alv-Gau&Ginsp85}
	Alvarez-Gaum\'e, L. \&  Ginsparg, P.``The structure of Gauge and Gravitational Anomalies" Ann. of Phys. 161, 423 (1985).

\bibitem{Bori&Ogie75} 
	Borisov, A.B. \& Ogievetskii I., ``Theory of Dynamical Affine and Conformal Symmetries as the theory of the Gravitational Field"
	Theor. Math. Phys. 21, 1179 (1975)
	[Teor. Mat. Fiz 21, 329 (1974)]

\bibitem{Sard02} 
	Sardanashvily, G., ``Classical Gauge Theory of Gravity "
	Theo. Math. Phys. 132 (2002) 1163-1171; 
	[Teor.Mat.Fiz. 132, 318 (2002)] (arXiv:gr-qc/0208054) 

\bibitem{Sard06} 
	Sardanashvily, G., ``Gauge Gravitation Theory from the Geometric Viewpoint"
 	Int. J. Geom. Methods Mod. Phys. V.3, N1 (2006) v, xx (arXiv:gr-qc/0512115) 

\bibitem{Sard11} 
	Sardanashvily, G., ``Classical Gauge Gravitational Theory",
	Int. J. Geom. Methods Mod. Phys. V8 (2011) 1869-1895 (arXiv:1110.1176, math-ph) 

\bibitem{Fult&Harr91}
 	Fulton, W. \& Harris, J. ``Representation Theory: a first course" Springer-Verlag 
	Graduated Texts in Mathematics.  Springer-Verlag New York, Berlin Heidelberg (1991). 

\bibitem{Kobayashi72}
	Kobayashi, S., ``Tranformation Groups in Differential Geometry" Springer, New York (1972).
 
\bibitem{DaSilvaSG06}
	Da Silva, A., "Lectures on Symplectic Geometry", Lecture Notes in Mathematics No. 1764, Springer-Verlag (2006).

\bibitem{Lecl06}
	Leclerc, M., "The Higgs sector of gravitational gauge theories" Ann. of Phys. 321, 708 (2006).

\bibitem{Low&Man02}
	Low, I \& Manohar, A., "Spontaneously Broken Spacetime Symmetries and Goldstone's Theorem", Phys. Rev. Lett. 88 (2002) 101602

\bibitem{SharpeE99}
	Sharpe, E.R., ``Discrete Torsion and Gerbes", arXive:hep-th/9909108 (1999)  

\bibitem{Sharpe02} 					
	Sharpe, E.R., ``Quotient Stacks and String Orbifolds", Nucl.Phys. B627 (2002) 445-505.

\bibitem{SharpeE03}
	Sharpe, E.R., ``Discrete Torsion", Phys.Rev.D68 (2003) 126003 

\bibitem{SharpeE01}
	Sharpe, E.R., ``String Orbifolds and Quotient Stacks", Nuc.Phys.B610 (2001) 595 
	
\bibitem{Olmos05} 
	Olmos, C.,  "A geometric proof of the Berger Holonomy Theorem", Ann. of Math. 161, 579 (2005).


\bibitem{Neu&Sij79}
	Ne'eman, Y. \& \v{S}ija\v{c}ki, Dj. Proc. Natl. Acad. Sci (USA) 76, 561 (1979).
	Ann. Phys. (N.Y.) 120, 292 (1979)

\bibitem{Morita97} Morita, S., ``Geometry of Differential Forms", AMS Translations of Mathematical Monographs, V201 (Japanese: 1997; English: 2001).

\bibitem{Palais59} Palais, R., ``Natural Operations on Differential Forms",  Trans.Amer.Math.Soc, 92, 125 (1959).

\bibitem{Kob&Nom56} Kobayashi, S. \& Nomizu, K., ``Foundations of Differential Geometry"  Wiley, N.Y. (1956).

\bibitem{McLa&Birk93} MacLane, S. \& Birkhoff, G. ``Algebra" Third Ed., A.M.S. Chelsea Publishing (1993).

\bibitem{Weib94} Weibel, C., ``An introduction to Homological Algebra", Cambridge Studies in Advanced Mathematics \#38, Cambridge Univ. Press (1994). 

\bibitem{Kapu&Thor13}
	Kapustin, A. \& Thorngren, R., ``Higer Symmetry and Gapped Phases of Gauge Theories", ArXiv:1309.4721 (2013).

\bibitem{Moer&Mrcu03} 
	Moerdijk, I. \& Mrcum, J., ``Introduction to Foliations and Lie Groupoids", Cambridge Studies in Advanced Mathematics \#91, Cambridge Univ. Press (2003). 

\bibitem{Moerd02} Moerdijk, I., ``Introduction to the Langague of Stacks and Gerbes", Third Lisbon Summer Lect. in Geometry (2002).

\bibitem{Lord&Gosw88} Lord, E. \& Goswami, P. ``Gauge theory of a group of diffeomorphisms. III. The fiber bundle description" J. Math. Phys. 29 (1), 258 (1988).

\bibitem{Neeb05}  Neeb, K.-H. ``Non-abelian extensions of infinite dimensional Lie groups (2005) arXiv:math/0504295 

\bibitem{Vakil12} Vakil, R. ``Math 216: Foundations of Algebraic Geometry" math216.wordpress.com, May 16, 2012 (draft) http://math.standford.edu/~vakil/216blog/

\bibitem{Neem&Thier80}
	Ne'eman, Y. \& Thierry-Mieg, J. ``Geometrical gauge theory of ghost and Goldstone fields and of ghost symmetries" Proc.Natl.Acad.Sci. USA 77, 720 
	(1980).

\bibitem{Freyd64}
	Freyd, Peter, ``Abelian Categories: An Introduction to the Theory of Funrctors"  Harper Series in Modern Mathematics, Harper \& Row 
	New Your, Evanston, and London (1964).  

\bibitem{Penrose84}
	Penrose, and Rindler, W., ``Spinors and Space-time" Vol.1, Cambridge Monographs in Mathematical Physics, Cambridge University Press (1984).

\bibitem{Sij98}
	\v{S}ija\v{c}ki, Dj. Acta Phys. Polo. B29, 1089 (1998)

\bibitem{tHooft76}
	`t Hooft, G., "Computation of the quantum effects due to a four-dimensional pseudoparticle", Phys.Rev.D 14, 3432 (1976). 

\bibitem{Sala&Stra69}
	Salam, A. \& Strathdee, J. ``Non-Linear Realizations. II. Conformal Symmetry" Phys.Rev. 184, 1760. 

\bibitem{Trau70} 					
	Trautman, A. (Fiber bundles associated with Spacetime) Rep. on Math. Phys. 1, 29 (1970).

\bibitem{VDBerg96}
	Van den Bergh, M. ``A translation principle for the four-dimensional Slkyanin algebras", J. Algebra 184 (1996) 435-490 					

\end{thebibliography}
}
\end{document}